\documentclass[10pt,sigconf,letterpaper]{acmart}

\usepackage{pdflscape}
\usepackage{xcolor}
\usepackage{comment}
\usepackage{float}
\usepackage{balance}
\usepackage{xspace}
\usepackage{subcaption}
\usepackage{multirow}
\usepackage{placeins}

\usepackage[font=small,format=plain,labelfont=bf,textfont=it]{caption}
\usepackage{placeins}
\usepackage{cleveref}
\hypersetup{citecolor = blue}
\usepackage[framemethod=TikZ]{mdframed}
\usepackage[utf8]{inputenc}
\graphicspath{{figures/}}

\makeatletter
\def\input@path{{sections/}{./}}
\makeatother

\crefformat{section}{\S#2\color{blue}#1#3} 
\crefformat{subsection}{\S#2\color{blue}#1#3}
\crefformat{subsubsection}{\S#2#1#3}

\setcopyright{none}
\renewcommand\footnotetextcopyrightpermission[1]{} 

\makeatletter
\def\subsubsection{\@startsection{subsubsection}{3}%
  \z@{.5\linespacing\@plus.7\linespacing}{.1\linespacing}%
    {\normalfont\itshape}}
    \makeatother

\pgfsetarrows{latex-latex}
\tikzset{%
  base/.style = {inner sep=5pt,
                 text centered,
                 thin,
                 font=\rmfamily},
  round/.style = {base,
                  rectangle,
                  rounded corners=1ex,
                  draw=black,
                  fill=gray!20,
                  minimum height=0.35in}
}

\newcommand{\teams}{Teams\xspace}
\newcommand{\meet}{Meet\xspace}
\newcommand{\zoom}{Zoom\xspace}

\newcommand{\zoombrowser}{Zoom-Chrome\xspace}
\newcommand{\teamsnative}{Teams-native\xspace}
\newcommand{\teamsbrowser}{Teams-Chrome\xspace}

\definecolor{ps0}{rgb}{0.97, 0.98, 1.00}
\definecolor{ps1}{rgb}{0.78, 0.86, 0.94}
\definecolor{ps2}{rgb}{0.42, 0.68, 0.84}
\definecolor{ps3}{rgb}{0.13, 0.44, 0.71}
\definecolor{ps4}{rgb}{0.03, 0.19, 0.42}

\usepackage[all=normal,wordspacing]{savetrees}

\title{Measuring the Performance and Network Utilization of Popular Video Conferencing Applications}

\author{Kyle MacMillan}
\affiliation{University of Chicago}

\author{Tarun Mangla}
\affiliation{
  \institution{University of Chicago}
}
\author{James Saxon}
\affiliation{%
  \institution{University of Chicago}
}
\author{Nick Feamster}
\affiliation{%
  \institution{University of Chicago}
}

\renewcommand{\paragraph}[1]{\vspace*{0.03in}\noindent\textbf{#1}}

\begin{document}

\begin{sloppypar}
\begin{abstract}
Video conferencing applications (VCAs) have become a critical Internet
application, even more so during the COVID-19 pandemic, as users worldwide now rely on them for
work, school, and telehealth. 
It is thus increasingly important to understand the resource requirements of different
VCAs and how they perform under different network conditions, including:
how much ``speed'' (upstream and downstream throughput) a VCA needs to support high quality of experience; how
VCAs perform under temporary reductions in available capacity; how they compete
with themselves, with each other, and with
other applications; and how usage modality (e.g., number of participants)
affects utilization.  We study three modern VCAs: Zoom, Google Meet,
and Microsoft Teams.   Answers to these questions differ substantially
depending on VCA.  First, the average utilization on an unconstrained
link varies between 0.8~Mbps and 1.9 Mbps.  Given temporary reduction of
capacity, some VCAs can take as long as 50 seconds to recover to steady state.  Differences in proprietary
congestion control algorithms also result in unfair bandwidth allocations: in
constrained bandwidth settings, one Zoom video conference can consume more than 75\%
of the available bandwidth when competing with another VCA (e.g., Meet, Teams).  For some VCAs, client utilization can
decrease as the number of participants increases, due to the reduced video
resolution of each participant's video stream given a larger number of
    participants. Finally, one participant's viewing mode
    (e.g., pinning a speaker) can affect the
upstream utilization of other participants.
\end{abstract}

\maketitle

\pagestyle{plain} 
\section{Introduction}\label{sec:intro}

Internet users around the world have become increasingly dependent on video
conferencing applications over the past several years, particularly as a
result of the COVID-19 pandemic, which has caused many users to rely almost
exclusively on these applications for remote work, education, healthcare,
social connections, and many other activities.  Some ISPs have reported as
much as a three-fold increase in video conferencing traffic during an
eight-month period in 2020 during the pandemic~\cite{bitag_report}. Our
increased reliance on these video conferencing applications~(VCAs) has highlighted
certain disparities, especially given that nearly 15 million
public school students across the United States during the pandemic lacked
Internet access for remote education~\cite{common_sense_report}. 

As cities and countries around the world moved to close this gap, many asked a
series of simple questions geared towards understanding the level of
connectivity that they needed to provide citizens to guarantee reliable,
high-quality Internet experience: {\em What is the baseline level of Internet
performance needed to support common video conferencing applications for the
activities people commonly use them for (e.g., education, remote work,
telehealth)?} This was the question, in fact, that our own city officials
asked us which motivated us to pursue this question in this paper. The question
has been brought into focus even more over recent months as the United Stated
Federal Communications Commission (FCC) continues to debate the levels of
Internet speed that should classify as a ``broadband'' service offering. The
current standard is 25~Mbps downstream/3~Mbps upstream. With the rise of video
conferencing applications, however, consumer Internet connections saw an
increase in upstream traffic utilization. Some policy advocacy papers have
claimed (without measurements) that the FCC should change its definition
of ``broadband'' to a symmetric 20~Mbps connection on account of these trends.
Moreover, because many users, especially in underserved regions, experience
frequent connectivity disruptions, a comparative analysis of VCAs under
dynamic (and degraded) network conditions can also shed light on the design
practices of VCAs and help identify best design practices.

In light of these discussions and questions from municipal, state, and federal
policymakers, we were motivated to explore the answers to questions concerning
how much network resources video conferencing applications required, how they
responded to network degradations and connectivity interruptions, how they
compete with each other and with other applications, and how these questions
vary depending on the modalities of use (e.g., gallery mode vs. speaker mode).
Similar questions have of course been studied in the
past~\cite{guha2005experimental, baset2004analysis, bonfiglio2008detailed,
bonfiglio2008tracking, xu2012video}, but the vast majority of these studies
are now at least a decade old, during which time both the VCAs themselves and
how we use them has changed dramatically. New VCAs such as Google Meet,
Microsoft Teams, and Zoom have emerged in the last few years alone. In light
of these new VCAs and recent drastic shifts in usage pattern
and our dependence on these applications, a re-appraisal of these questions is
timely.

We study the following questions for three popular, modern video conferencing
applications:
\begin{enumerate}
    \itemsep=-1pt
    \item What is the network utilization of common video conferencing
        applications?
    \item What speeds do these VCAs require to deliver good quality of
        experience to users?
    \item How do VCAs respond to temporary disruptions to connectivity?
        and how quickly do they recover when connectivity is restored?
    \item How do VCAs respond in the presence of competing traffic? How do
        they compete with themselves, with other VCAs, and with other
        applications?
    \item How do different usage modalities (number of participants, speaker
        vs. gallery viewing) affect network utilization?
\end{enumerate}
\noindent
We explore these questions by performing controlled laboratory experiments
with emulated network conditions across a collection of clients, collecting
application performance data using provided application APIs. We collect
application data under a variety of network conditions and call modalities,
including different numbers of participants and viewing modes. Many of the answers
to these questions surprised us; as we will discover,
sometimes the answers depend quite a lot on the particular video
conferencing application. Table~\ref{tab:results} summarizes our main findings
and where they can be found in the paper. In particular, we found significant
differences in network utilization, encoding strategy, and recovery time after
network disruption. We also discovered that VCAs share available network
resources with other applications differently. We confirm earlier findings 
\cite{cable-labs-vca}, that one participant's choice of how to interact 
in the video conference can affect the network utilization of {\em other} participants.

Our findings have broad implications, for network management, as well as for
policy. For network management, our findings concerning network utilization
and performance under various network conditions have implications for network
provisioning, as access ISPs seek to provision home networks to achieve good
performance for consumers. From a policy perspective, our results are
particularly important: questions about the throughput needed to support
quality video conferencing, especially in multi-user households, was a
question from city government officials that motivated this paper in the first
place. Looking ahead, as city, state, and federal governments both subsidize
broadband connectivity and build out new infrastructure, understanding how
these increasingly popular video conferencing applications consume and share
network resources is of critical importance.

\begin{table}[t]
    \begin{small}
    \begin{tabular}{|p{3.25in}|}
 \hline 
    The average network utilization on an unconstrained link ranges from 0.8
        Mbps to 1.9 Mbps (\cref{subsec:network_utilization}).\\ \\
    Despite using the same WebRTC API, Meet and Teams browser clients differ significantly in how they encode video (e.g., frames per second, resolution) under reduced link capacity (\cref{subsec:application_performance}).
\\ \\
    Recovery times following transient drops in bandwidth are quite different
        and depend on both VCA's congestion control mechanism and the
        direction of the drop, i.e. uplink or downlink. However, all VCAs take
        at least 20 seconds to recover from severe uplink drops to 0.25 Mbps
        (\cref{sec:interruption}).\\\\
    Differences in proprietary congestion control also create unfair bandwidth
        allocations in constrained bandwidth settings. Zoom and Meet can
        consume over $75\%$ of the available downlink bandwidth when competing
        with Teams or a TCP flow (\cref{sec:competition}).\\\\
    Each participant's video layout impacts its own and others' network utilization. Pinning a user's video to the screen (speaker mode) leads to an increase in the user's uplink utilization (\cref{sec:usage_modality}). 
    \\ \hline
\end{tabular}
    \end{small}
    \caption{Main results, and where they can be found in the paper.\label{tab:results}}
\end{table}

\section{Background \& Experiment Setup}
\label{sec:background}

We provide a brief background on video conferencing transport, technology, and
architecture before turning to our experiment setup.

\subsection{Video Conferencing Applications}

Most modern VCAs use Real-time Transport Protocol (RTP)~\cite{schulzrinne1996rtp,
schulzrinne2003rfc3550} or its variants~\cite{baugher2004secure, zoom_rtp} to
transmit media content. The audio and video data is generally transmitted over
separate RTP connections. RTP also uses two other protocols in conjunction,
Session Initiation Protocol (SIP)~\cite{rosenberg2002sip} to establish
connection between clients and RTP Control Protocol
(RTCP)~\cite{schulzrinne2003rfc3550} to share performance statistics and
control information during a call. Despite using well-known protocols, VCAs
can differ from each other significantly across the following dimensions:

\begin{itemize}
    \itemsep=-1pt
    \item \textbf{Network mechanisms}: RTP and the associated protocols are
        implemented in the application-layer. Thus, the specific
        implementation of the protocols can vary across VCAs. For instance,
        Zoom reportedly uses a custom extension of RTP~\cite{zoom_rtp}).
        Furthermore, the key network functions, i.e., congestion control and
        error recovery, can also be different and proprietary. 

    \item \textbf{Application-layer parameters}: These include media encoding
        standards and default bitrates. More recent video codecs (e.g., VP9,
        H.264) can encode at the same video quality with fewer bytes when
        compared to older codec (e.g., VP8), albeit with a higher compute
        overhead~\cite{bienik2016performance}. 
    
    \item \textbf{Streaming architecture}: VCAs can either choose to use
        direct connections or use relay servers. Centralized servers are
        almost always used for multi-point calls to combine data from multiple
        users. VCAs can differ in the exact strategies of data combination as
        well as the geographic footprint of their servers. For instance, Zoom
        rapidly expanded its server infrastructure to support increased call
        volume because of COVID-19~\cite{liu2020characterizing}. 

\end{itemize}
\noindent
Differences across one or more of these factors can lead to different VCA
network utilization and performance. In this paper, we aim to dig deeper into
some of these differences for a subset of VCAs.

\subsection{Experiment Setup} 
\label{subsec:setup}

\paragraph{Video Conferencing Applications}:
In this paper, we study three popular VCAs---\zoom, Google \meet, and Microsoft \teams.
These VCAs have been used extensively worldwide over the past year, especially
in enterprises and educational institutions~\cite{vca_share}.  Teams and Zoom
provide desktop applications as well as browser clients, whereas Meet is ``native"
in the Chrome. Most of our tests are conducted using the desktop
applications for Teams and Zoom, and using the Google Chrome browser for Meet.
In-browser tests for Teams and Zoom are specified by \textit{\teamsbrowser}
and \textit{\zoombrowser}, respectively. We use Chrome (Meet) version
89.0.4389, Zoom client version 5.6.1, and Teams client version 1.4.00.7556.

\paragraph{Laboratory Environment}:
We conduct experiments in a controlled environment.  We begin by describing our experimental setup for a 2-party call. We use two identical laptops, referred to as C1 and C2, representing the two VCA clients. Each laptop is a Dell Latitude 3300 with a screen resolution of 1366 $\times$
768 pixels and running Ubuntu 20.04.1. The laptops have a wired connection to
a Turris Omnia 2020 router and access a dedicated 1~Gbps symmetric link
to the Internet.  Each experiment consists of a call between C1 and C2 under a
pre-specified network bandwidth profile and VCA. The bandwidth profile is
emulated by shaping the link between C1 and the router using traffic control
(\texttt{tc}). A pre-recorded talking-head video\footnote{It would be inappropriate to use the 
device webcam as the video, because without movement, VCAs compress the video
and ultimately send at a much lower rate than during a normal call.} with a
resolution of 1280 $\times$ 720 is used as the video source for the call, using \texttt{ffmpeg}.
This is done to both replicate a real video call and ensure consistency across
experiments. All experiments are conducted with the laptop lid open and the
application window maximized. 

\begin{figure*}[t!]
\begin{subfigure}[t]{0.33\textwidth}
    \centering
    \includegraphics[width=\textwidth,keepaspectratio]{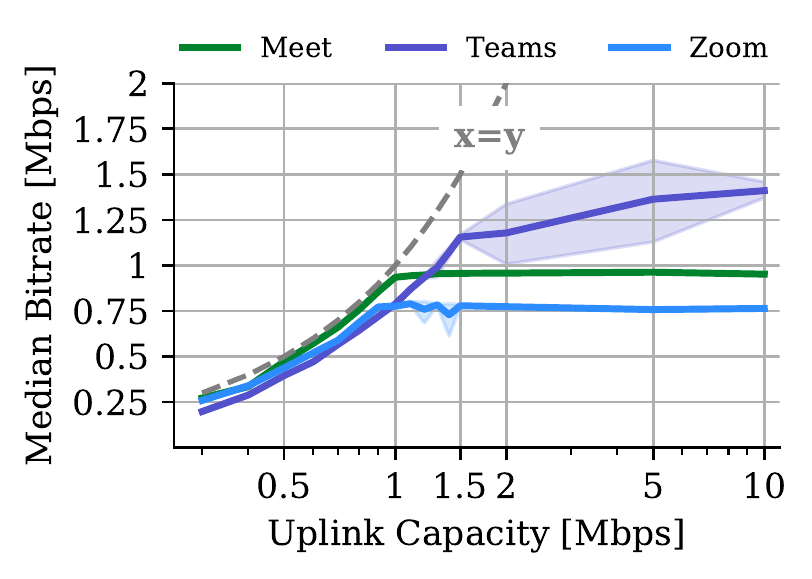}
    \caption{Uplink bandwidth vs network bitrate 
    }
	\label{subfig:uplink_bitrate}
\end{subfigure}\hfill
\begin{subfigure}[t]{0.33\textwidth}
\centering
    \includegraphics[width=\textwidth,keepaspectratio]{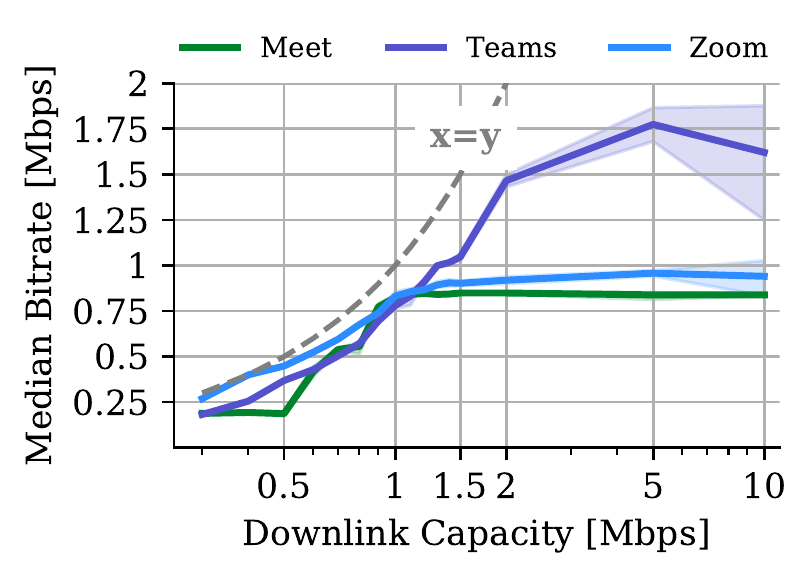}
    \caption{Downlink bandwidth vs network bitrate}
	\label{subfig:downlink_bitrate}
\end{subfigure} \hfill
\begin{subfigure}[t]{0.33\textwidth}
\centering
    \includegraphics[width=\textwidth,keepaspectratio]{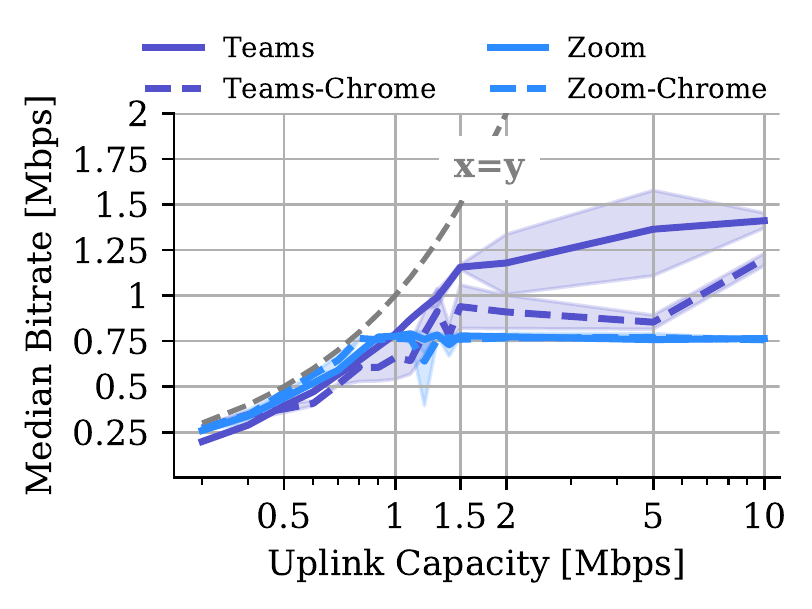}
    \caption{Impact of VCA platform 
    }
	\label{subfig:uplink_browser}
\end{subfigure} 
\vspace{-1em}
\caption{Utilization under different shaping levels. The bands represent 90\% confidence intervals.}
\label{fig:static}
\end{figure*}

\paragraph{Automating Experiments}: To conduct experiments at scale, we
automated the entire in-call process. We take several steps to recreate the
in-call process.  We use the Python PyAutoGUI package~\cite{pyautogui} to
automate joining and leaving calls. The package enables to programmatically
control the keyboard and the mouse by specifying coordinates or visual
elements on the screen. For \zoombrowser, we encountered CAPTCHA before
joining a call on the default browser. Using the Selenium-based Chrome
browser~\cite{selenium}, however, enabled us to bypass the CAPTCHA. Note the
experiments using Selenium are run exactly as it would be in default Chrome
browser. The workflow is controlled from C1, with TCP sockets used to
coordinate between C1 and C2.  We modify this setup slightly for subsequent
experiments (e.g., multi-party calls); those slight modifications are
described in the respective sections.

\section{Static Network Conditions}\label{sec:static}

In this section, we study the effect of network capacity on VCA performance.

\paragraph{Method}: We conduct a series of experiments, each consisting of a
2.5-minute call between two clients, C1 and C2 (as described in
Section~\ref{subsec:setup}), under a specific shaping level. We conduct two
sets of experiments, shaping first the uplink and then the downlink.  We
constrain throughput to \{$0.3, 0.4, \dots, 1.4, 1.5, 2, 5, 10$\}~Mbps (in
each of the upstream and downstream directions). For each condition, we perform five 2.5-minute experiments; our
data show that due to relatively low variance in most observations, this
number of experiments is sufficient to achieve statistically significant
results.  For \zoom and \teams clients, we perform measurements both for
native clients as well as for browser-based clients, referred to as
\zoombrowser and \teamsbrowser, respectively. 



\subsection{Network Utilization}
\label{subsec:network_utilization}

\paragraph{Constrained upstream utilization}: Figure~\ref{subfig:uplink_bitrate} shows the
median sent network bitrate for different uplink capacities. We observe differences in upstream network
utilization among VCAs given the same available uplink network capacity. In
the case of a 10~Mbps uplink, for example, the average upstream utilization for
\teamsnative is $1.44$ Mbps whereas it is only $0.95$ for \meet and $0.77$
Mbps for \zoom. All three VCAs utilize the uplink efficiently (above 85\%)
when that link is constrained (0.8 Mbps or lower), although \meet's
bitrate is slightly higher than the other VCAs.  

\begin{table}[t]
\centering
\begin{tabular}{|c|c|c|}
\hline
\multirow{2}{*}{\textbf{VCA}} & 
    \multicolumn{2}{c|}{{\bf Utilization (Mbps)}} \\ 
    \cline{2-3} 
                              & Upstream                   & Downstream                  \\ \hline
Meet                          & 0.95                     & 0.84                      \\ 
Teams                         & 1.40                      & 1.86                      \\ 
Zoom                          & 0.78                     & 0.95                      \\ \hline
\end{tabular}
\caption{Unconstrained network utilization.}
\vspace{-1.5em}
\label{tab:vca_static}
\end{table}

\paragraph{Constrained downstream throughput}: We explored the effect of constrained
downstream capacity
on VCAs' network utilization. Figure~\ref{subfig:downlink_bitrate} shows this
result.
As with a constrained uplink, the VCAs differ in terms of their downlink
utilization under unconstrained link. The unconstrained downlink utilization
differs from unconstrained upstream utilization, as shown in
Table~\ref{tab:vca_static}. To better understand this phenomenon, we analyzed the traffic
captured from both clients. 

In the case of \teams, we found that the sent traffic from C1
is almost same as the received traffic at C2, and vice versa. 
We suspect the small
differences in upstream and downstream utilization may largely be due to the
variability in utilization across experiments in \teams, as is
also evident in the larger confidence intervals as compared to \zoom and \meet. 

In contrast, \zoom's utilization patterns were asymmetric: we found an
asymmetry in sent and received data rates at both clients. For instance, in a
single instance with 10~Mbps downlink shaping, C2 sent a median 0.85~Mbps and
C1 received a median 1.10 Mbps. Investigating this asymmetry further, we
discovered an interesting phenomenon: We found that \zoom uses a relay server
instead of direct communication. Additionally, related work by Nistico et
al.~\cite{nistico2020comparative} suggests that \zoom uses Forward Error
Correction (FEC) for error recovery. This is further supported by a related
patent from \zoom itself, which talks about a methodology to generate FEC data
at the server~\cite{liu2019error}. We suspect that the extra data may thus
correspond to FEC added by the relay server, leading to asymmetric upstream and
downstream utilization.  

\begin{figure*}[t]
    \begin{subfigure}[t]{0.33\textwidth}
    		\centering
        \includegraphics[width=\textwidth,keepaspectratio]{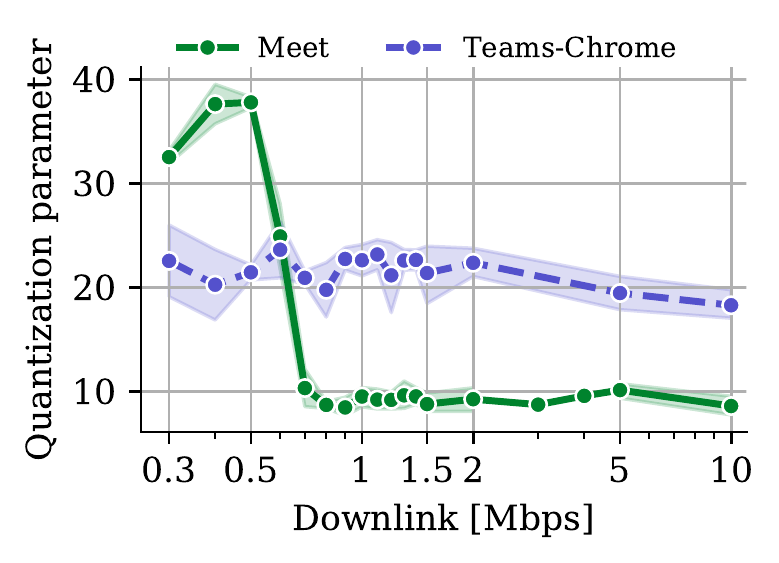}
        \caption{Downlink - Quantization parameter}
 		\label{subfig:downlink_video_qp}
    \end{subfigure}%
    \hfill
	\begin{subfigure}[t]{0.33\textwidth}
        \centering
        \includegraphics[width=\textwidth]{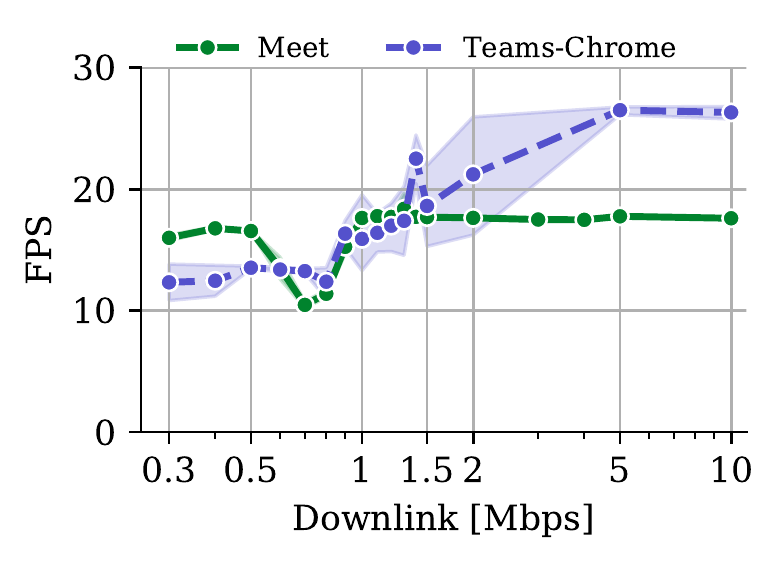}
    \caption{Downlink - Frames per second.}
    \label{subfig:downlink_frames_per_second}
    \end{subfigure}%
    \hfill
	\begin{subfigure}[t]{0.33\textwidth}
        \centering
        \includegraphics[width=\textwidth]{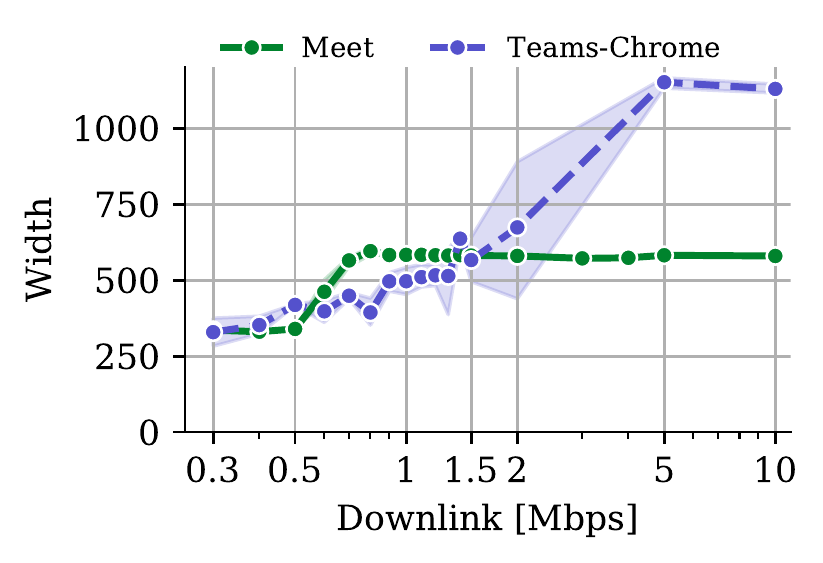}
    \caption{Downlink - Frame width.}
    \label{subfig:downlink_frame_width}
    \end{subfigure}
    \newline
        \begin{subfigure}[t]{0.33\textwidth}
    		\centering
        \includegraphics[width=\textwidth,keepaspectratio]{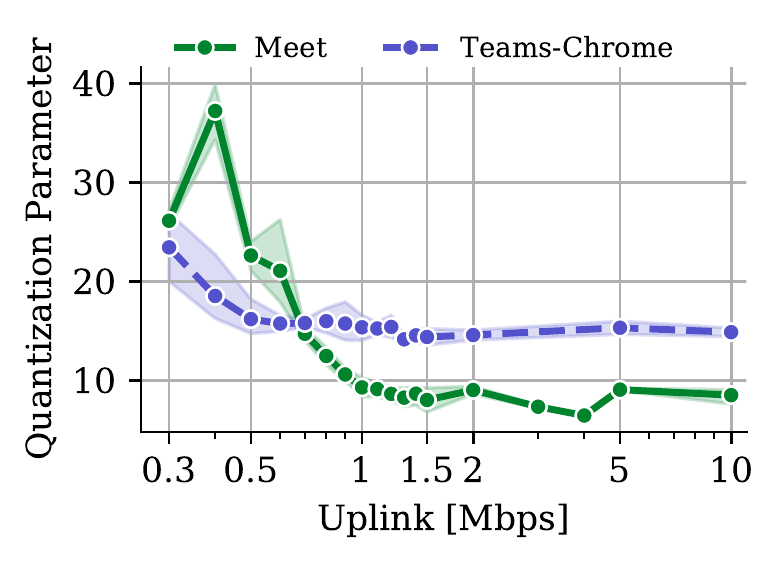}
        \caption{Uplink - Quantization parameter.}
 		\label{subfig:uplink_video_qp}
    \end{subfigure}%
    \hfill
	\begin{subfigure}[t]{0.33\textwidth}
        \centering
        \includegraphics[width=\textwidth]{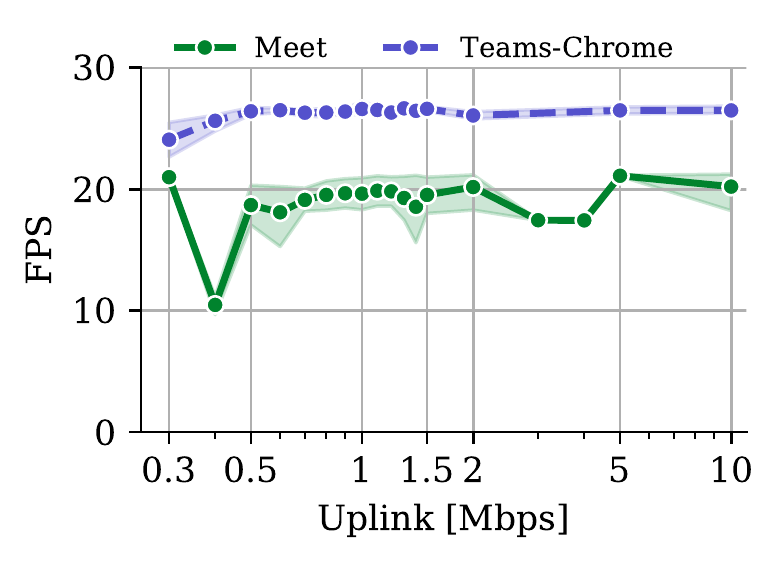}
    \caption{Uplink - Frames per second.}
    \label{subfig:uplink_frames_per_second}
    \end{subfigure}%
    \hfill
	\begin{subfigure}[t]{0.33\textwidth}   
        \centering
        \includegraphics[width=\textwidth]{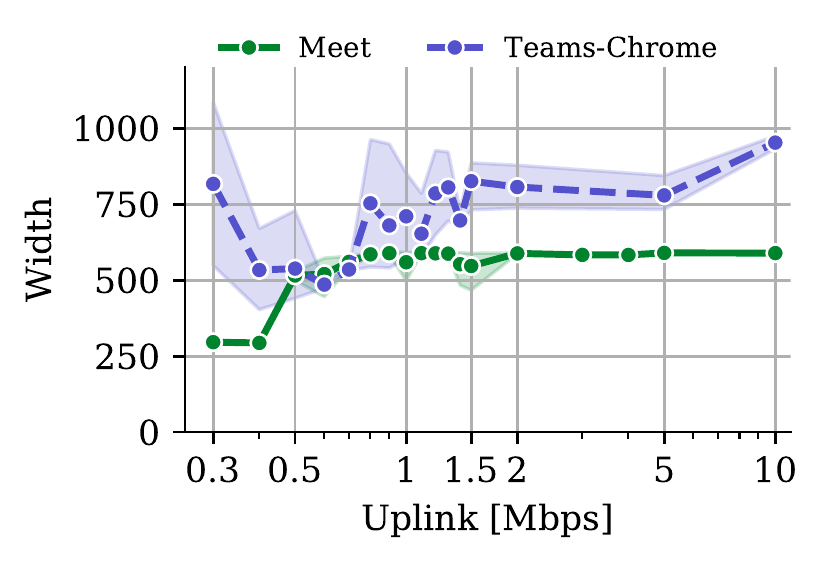}
    \caption{Uplink - Frame width.}
    \label{subfig:uplink_frame_width}
    \end{subfigure}
	\caption{Video encoding parameters with 90\% confidence intervals under
    downstream and upstream throughput constraints.}
    \vspace{-1em}
	\label{fig:video_qual}
\end{figure*}

In addition to the asymmetric unconstrained utilization, \meet exhibits
markedly different behavior with a constrained downlink.
Specifically, the
network utilization with constrained downstream throughput (< 0.8 Mbps) is
only 39--70\%
(Figure~\ref{subfig:downlink_bitrate}), while it is more than $90\%$ in the
case of a constrained uplink (Figure~\ref{subfig:uplink_bitrate}). Upon
further exploration, we discovered that
\meet also uses a relay server, as well as
\textit{simulcast}, wherein the sender (C2) transmits multiple copies of the
video to the server, each at a different quality
level~\cite{nistico2020comparative}. The server then relays one of the quality
streams to C1 depending on the inferred available capacity of the server-C1
link. We observe two simultaneous video streams in our experiments, one at
320x180 and other at 640x360 resolution. When downstream throughput is
constrained, the relay
server cannot switch to a higher quality video and keeps sending at low
quality bitrate. This explains why \meet's network utilization at 0.5 Mbps is
only 0.19~Mbps, almost similar to its utilization at 0.3 Mbps.  The use of
simulcast also explains the higher upstream utilization as compared to
downstream utilization. 

\paragraph{Browser vs. native client utilization}:
Figure~\ref{subfig:uplink_browser} compares the upstream utilization of \zoom
and \teams between their respective native and Chrome clients. \zoom's
utilization is similar across the native and browser-based platforms; in
contrast, we find significant difference between \teams native and browser
client. When uplink capacity is shaped to 1~Mbps, the \teams-native client
uses 0.84 Mbps, whereas \teamsbrowser uses only 0.61 Mbps. We found a
similar difference between \teams-native and \teamsbrowser when downstream
throughput is constrained. 

\begin{figure}[t]
    \centering
    \begin{subfigure}[t]{0.4\textwidth}      
        \includegraphics[width=\textwidth,keepaspectratio]{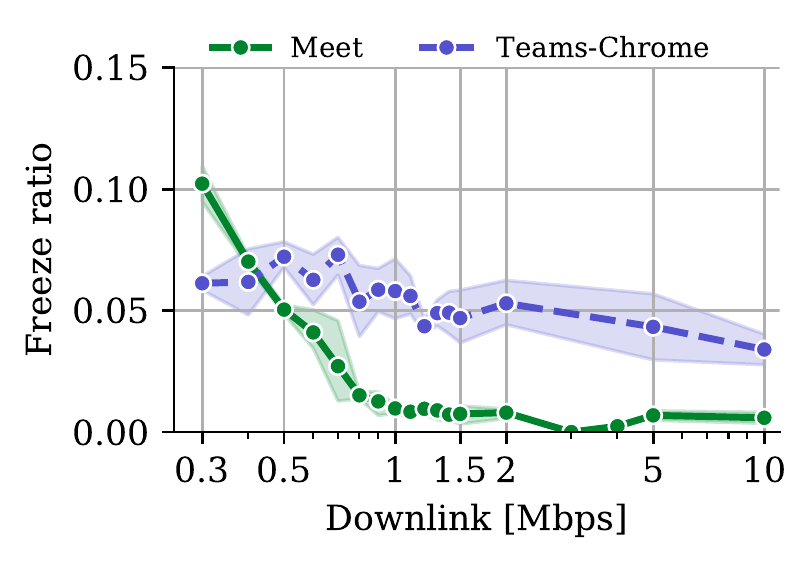}
        \vspace{-1em}
        \caption{Downstream: Freeze ratio.}
 		\label{subfig:downlink_freeze_ratio}
    \end{subfigure}
	\begin{subfigure}[t]{0.4\textwidth}   
        \centering
        \includegraphics[width=\textwidth]{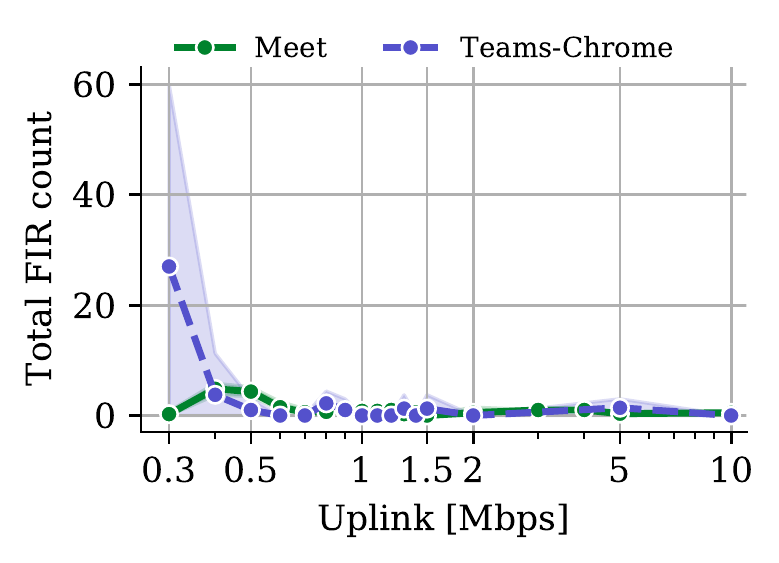}
    \vspace{-1em}
    \caption{Full Intra Request (FIR) count.}
    \label{subfig:uplink_fir}
    \end{subfigure}%
    \vspace{-1em}
	\caption{Video freeze under downstream and upstream throughput constraints. The bands represent 90\% confidence intervals.
	}
	\label{fig:video_freeze}
\end{figure}

\subsection{Application Performance}
\label{subsec:application_performance}

Next, we explore how video conference application performance varies depending
on network capacity.  To do so, we rely on the WebRTC stats API
available in Google Chrome to obtain application metrics for \teamsbrowser and
\meet~\cite{webrtc_stats}. We could not obtain the same statistics for
\zoombrowser as it uses DataChannels instead of RTP MediaStream in WebRTC to
transmit media. DataChannels statistics lack any video quality metrics and
mostly contain data volume information. Obtaining application metrics is
challenging for native clients. The Zoom API does provide limited
application performance (e.g., video resolution, FPS) at a per-minute
granularity~\cite{zoom_qos} for the native client, but this granularity is insufficient to
observe short-term quality fluctuations.  We thus limit our analysis in this
section to only \meet and \teamsbrowser. We focus on a subset of metrics
available from WebRTC Stats API that relate to video \textit{quality} and
\textit{freezes}. These metrics are available at a per-second granularity
during each call. 

\paragraph{Video quality}: VCAs adapt the video quality by adjusting the
encoding parameters to achieve a target bitrate estimate provided by the
transport. Ideally, VCAs can adjust one or more of the following three
parameters: 
\begin{itemize}
    \itemsep=-1pt
    \item \emph{frames per second} (FPS), 
    \item \textit{quantization parameter} used in video compression, and  
    \item \textit{video resolution} indicated as the number of pixels in each dimension. 
\end{itemize}
\noindent
We can obtain all three
parameters from the WebRTC stats. In our analysis, we present a single
dimension of the resolution, namely, its \textit{width}. 

We first look at these parameters for different downstream 
capacities
(see~\Cref{subfig:downlink_video_qp,subfig:downlink_frames_per_second,subfig:downlink_frame_width}).
\teamsbrowser simultaneously degrades all three video quality parameters as
downstream capacity decreases. In the case of \teamsbrowser, we also observe
variable behavior across multiple experiments under the same network capacity,
as shown by the $90\%$ confidence interval bands in the plot. \meet, on the
other hand, behaves in a more consistent fashion: it adjusts parameters
differently based on capacity. Within the  0.7--1~Mbps range, \meet adapts the
bitrate primarily by adapting {frames per second}, while keeping the
{quantization parameter} and {frame width} similar to the quality levels in
the absence of degradation. As capacity is further constrained, however,
from 0.5--0.7~Mbps, both the {frame width} and {quantization parameter}
degrade, whereas there is a simultaneous {\em increase} in the frames per
second. Upon closer inspection of the per-second statistics in this operating
range, we find that the relay server may switch to a lower quality copy of the
stream it received via {simulcast}. \meet does not appear to reduce the
bitrate any further by reducing the FPS for the low-quality
stream---specifically, we observed a consistent frame rate even at rates of
less than 0.5~Mbps It is not clear why the quantization parameter reduces
from 38 to 33 at 0.3 Mbps in \meet.

We next analyze the effect of uplink capacity constraints on encoding
parameters
(see~\Cref{subfig:uplink_video_qp,subfig:uplink_frames_per_second,subfig:uplink_frame_width}).
\teams adapts to constrained throughput settings mainly by increasing the
quantization parameter and reducing the frame width, while keeping the FPS
almost constant. To our surprise, we found that the frame width, {\em
increases} as uplink capacity is reduced to 0.3~Mbps.
There seems to be no particularly good explanation for this effect, and (as we
discuss in the next paragraph) it also results in video freezes, suggesting a
poor design decision or implementation bug.  \meet follows a similar trend by
mostly increasing the quantization parameter until the upstream bandwidth
decreases to 0.5 Mbps. At 0.4 Mbps, it also reduces frame width and the FPS. 

\paragraph{Video freezes}: We also analyze the effect of constrained settings
on video freezes. When constraining downlink capacity, we directly obtain the freeze duration
from WebRTC stats. A freeze is assumed to occur if the frame
inter-arrival $>$ max (3$\delta$, $\delta$ + $150 ms$), where $\delta$ is the
average frame duration. We normalize the freeze duration with the total call
duration to obtain freeze ratio.  Figure~\ref{subfig:downlink_freeze_ratio}
shows the freeze ratio under different downstream capacities. The freeze ratio
increases as the downlink bandwidth degrades. \meet has higher freeze ratio than \teamsbrowser with $10\%$ freeze ratio at 0.3~Mbps. Interestingly,
\teamsbrowser incurs freezes (a $3.6\%$ freeze ratio) {\em even in the absence
of throughput constraints}, suggesting again implementation problems or poor
design choices.

When the uplink is constrained, we could not obtain freeze statistics
from C1's logs, because the WebRTC stats provide freeze statistics only for
the received video stream.  Instead, we analyze the total count of Full Intra
Requests (FIR) for the upstream video data. An FIR is sent if the receiver
cannot decode the video or falls behind, likely due to frame losses.  A low
FIR count does not rule out freezes on the receiver, but a high count does
indicate that video freezes are occurring.  Figure~\ref{subfig:uplink_fir}
shows that the FIR count is particularly high for \teamsbrowser at uplink
capacity below 0.5~Mbps. A high FIR count in \teamsbrowser may be triggered
due to the sender sending high-resolution video, as shown in
Figure~\ref{subfig:uplink_frame_width}.

\vspace{3pt}
\begin{mdframed}[roundcorner=5pt, backgroundcolor=black!10]
\paragraph{Takeaways}: VCA performance under same network conditions varies
    among tested VCAs. Further, the VCAs' average upstream utilization on an unconstrained link has implications for broadband policy. The FCC currently recommends a 25/3 Mbps minimum connection. Such a connection may not suffice even for two simultaneous video calls. 
\end{mdframed}

\section{Network Disruptions}
\label{sec:interruption}

VCAs must handle network disruptions; they may do so in different ways.
Designing to cope with disruptions, such as interruptions to network
connectivity, has become even more pertinent over the past year as users have
increasingly relied on broadband Internet access to use VCAs, which can
sometimes experience connectivity disruptions.  In addition to experiencing
periodic outages, home networks are susceptible to disruptions caused by
temporary congestion along the end-to-end path.
In this section, we aim to understand how VCAs respond to the types of
disruptions that home Internet users might sometimes experience.

\paragraph{Method}:
We analyze how
VCAs respond to temporary network disruptions during the call by introducing
transient capacity reductions. Using the same setup as in
the previous experiment, we initiate a five-minute VCA session between
two clients, C1 and C2, both of which are connected to the Internet via a
1~Gbps link. One minute after initiating the call, we reduce the capacity
between C1 and the router for 30 seconds, before reverting back to the
unconstrained 1~Gbps link. We repeat each experiment four times.
We conduct two sets of experiments: first disrupting the uplink, and then
disrupting the downlink. We reduce capacity to the following
levels: {0.25, 0.5, 0.75, 1.0} Mbps. (We do not consider disruptions in capacity
to levels of more than 1~Mbps because both Zoom and Meet's average utilization
is below 1~Mbps.)

\begin{figure}[t!]
\centering
\begin{subfigure}[t]{.5\textwidth}
    \centering
    \includegraphics[width=.7\textwidth,keepaspectratio]{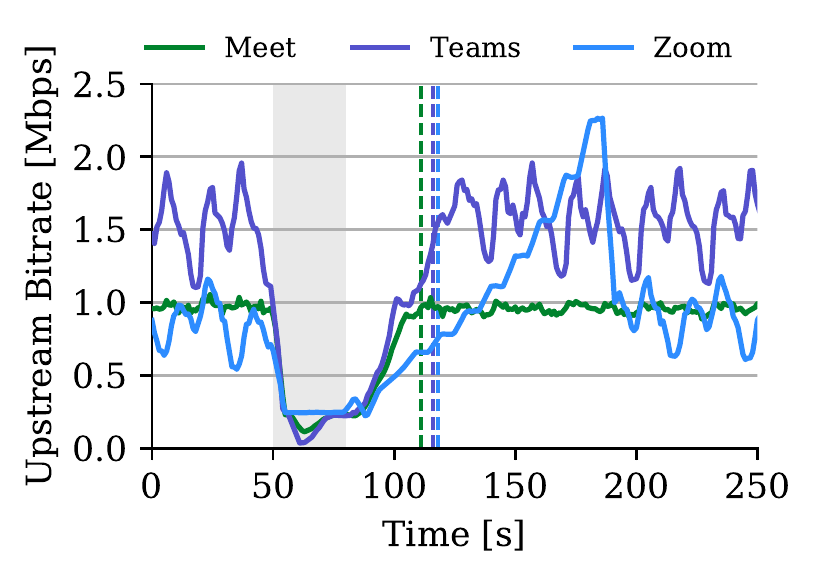}
    \caption{Average upstream bitrate over time. Grey region indicates period where uplink capacity is constrained to 0.25 Mbps. Vertical dotted lines indicate when the upstream bitrate has returned to the average bitrate before disruption.}
    \label{fig:ts_upld}
\end{subfigure}\hfill
\begin{subfigure}[t]{.5\textwidth}
      \centering
    \includegraphics[width=.7\textwidth,keepaspectratio]{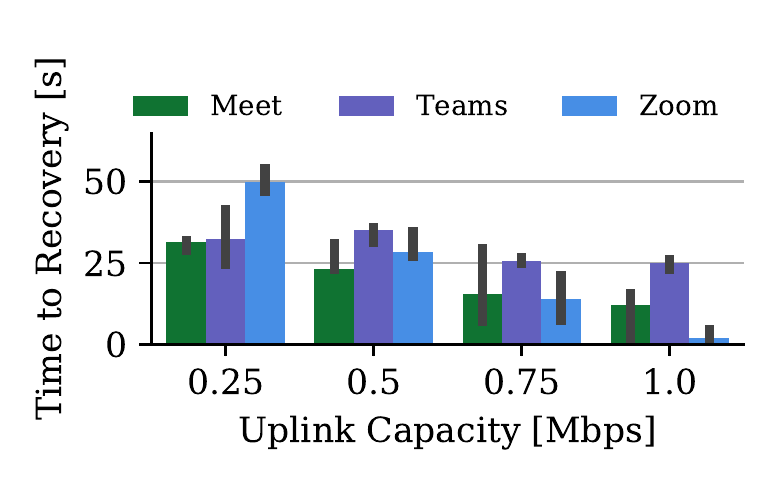}
    \caption{Time to recover from disruption to a given uplink capacity}
    \label{fig:TTR_upld}
\end{subfigure}
\caption{VCA response to a 30-second reduction in uplink capacity.}
\label{fig:interrupt-upld}
\end{figure}

\subsection{Uplink Disruptions}

Figure \ref{fig:ts_upld} shows VCA upstream bitrate over the course of a
call. Following a disruption, both the time to recovery and the
characteristics of that recovery
differ across VCAs. We quantify the time for
VCA utilization to return to normal by defining a 
\textit{time to recovery} (TTR) metric. We define TTR as the time between when the
interruption ends and when the five-second rolling median bitrate reaches the
median bitrate before interruption, also referred to as nominal bitrate. 

\paragraph{Time to Recovery}: Figure \ref{fig:TTR_upld} shows how the extent of
the disruption to uplink connectivity affects each VCA's time to recovery.
The more severe the capacity reduction, the longer the VCAs need to recover.

\teams takes longer to recover even at less severe disruptions for two
reasons: (1)~the nominal bitrate of Teams is higher than Meet and Teams,
(2)~Teams increases the upstream bitrate slowly immediately after the
interruption before increasing quickly back to normal (as shown in
Figure~\ref{fig:ts_upld}). Meet also observes a similar recovery pattern when
the disruption drops capacity to 0.25~Mbps. However, it recovers much faster
for the case of less severe disruptions, mostly because its nominal bitrate is
around 0.96 Mbps. 

\paragraph{Recovery Patterns}: While Meet and Teams follow a more rapid trend,
Zoom's recovery is different: It takes the longest time to recover under
severe disruptions. According to Figure \ref{fig:ts_upld}, Zoom follows a
stepwise recovery, with an almost-linear increase immediately after
interruption. It then enters a periodic-probing phase where it increases
the sending rate, stays at it for sometime before increasing it again. The
probing phase continues well above its nominal bitrate, before finally
reducing the bitrate.  Zoom does not return to a steady-state sending pattern
until two minutes after the disruption, in the meantime sending at much higher
rates. At first glance, such probing might appear to be bad design as it could
introduce additional packet loss and delay harming both Zoom's own performance
and the performance of other applications.  We suspect, however, that Zoom may
be using redundant FEC packets to gauge capacity similar to the
FBRA congestion control proposed by Nagy et al.~\cite{nagy2014congestion}.
Thus, even if such behavior induces packet loss, the user performance may not
suffer. Nevertheless, this inefficient use of the uplink, however, could
disturb other applications on the same link, leading to a poor quality of
experience for competing applications. 

\begin{figure}[t!]
 \centering
\begin{subfigure}[t]{.5\textwidth}
   \centering
    \includegraphics[width=.7\textwidth,keepaspectratio]{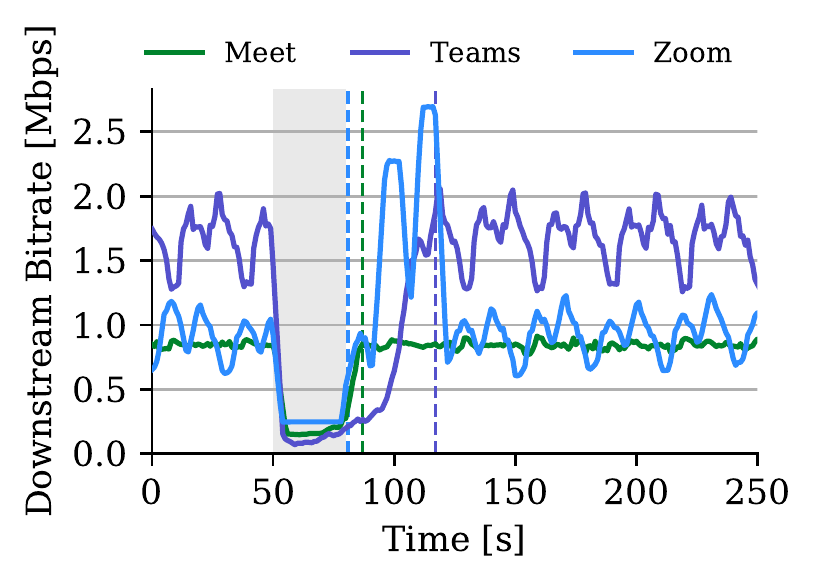}
    \caption{Average downlink bitrate over time. Grey region indicates period where downlink capacity is constrained to 0.25 Mbps. Vertical dotted lines indicate when the downstream bitrate has returned to the average bitrate before disruption.}
    \label{fig:ts-dnld}
\end{subfigure}
\begin{subfigure}[t]{.5\textwidth}
  \centering
    \includegraphics[width=.7\textwidth,keepaspectratio]{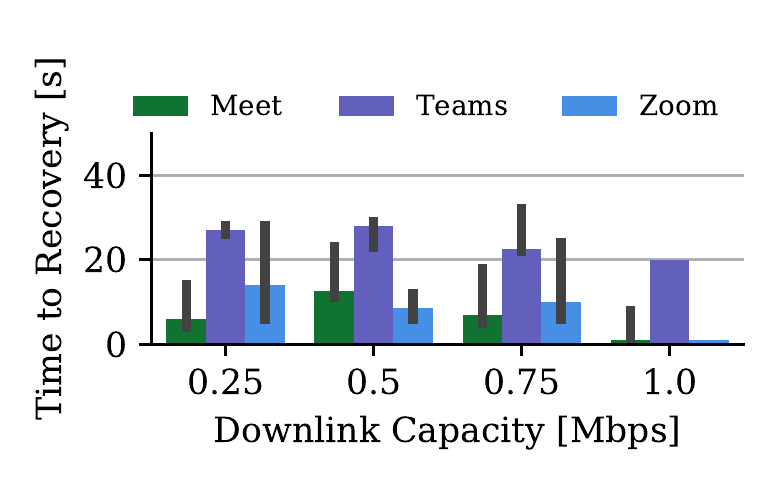}
    \caption{Time to recover from disruption to a given downlink capacity.}
    \label{fig:TTR_dnld}
\end{subfigure}
\caption{Response to a 30-second reduction in downlink capacity.}
\label{fig:interrupt-dnld}
\end{figure}

\subsection{Downlink Disruptions}

Figure~\ref{fig:TTR_dnld} shows that Teams recovers more slowly than Meet and
Zoom, always taking at least 20 seconds longer to return to the average rate,
regardless of the magnitude of the interruption. Furthermore, Meet and Zoom
recover much more quickly in this case compared to uplink interruptions. This
can be explained by the way each VCA sends video. For all three VCAs, the
clients communicate through an intermediary server. Thus, the recovery times
depend on the congestion control behavior at the server, as well. 

As mentioned in Section~\ref{sec:static}, Meet uses an encoding technique
called simulcast, where the client encodes the same video at different quality
levels and sends the encoded videos to an intermediate server. The server then
sends the appropriate version based on the perceived downlink capacity at the
receiver. The server can quickly switch between versions based on downstream
capacity. This quick recovery is shown in
Figure~\ref{fig:interrupt-dnld}: Meet returns to its average rate in under ten
seconds, regardless the severity of the interruption.

Similarly, Zoom uses \textit{scalable video coding} when transmitting
video~\cite{zoom_encoding}. Instead of sending several versions of varying
quality, Zoom sends hierarchical encoding layers that, when combined,
produce a high quality video. This allows C2 to send uninterrupted
even when C1's downlink capacity decreases. The server can then recover faster
by sending additional layers once the network conditions improve.  

\begin{figure}[t]
    \centering
    \includegraphics[width=0.35\textwidth,keepaspectratio]{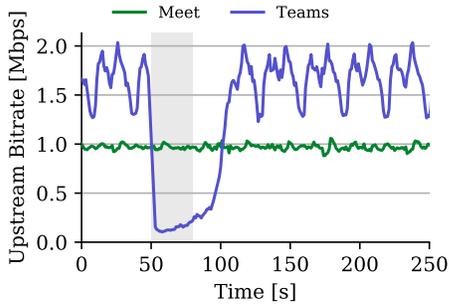}
    \caption{Client 2 (C2) upstream bitrate. Grey region indicates when C1's downlink capacity is reduced to 0.25 Mbps}
    \label{fig:interrupt-sender}
\end{figure}

\begin{figure}[]
   \centering
    \includegraphics[width=0.5\textwidth,keepaspectratio]{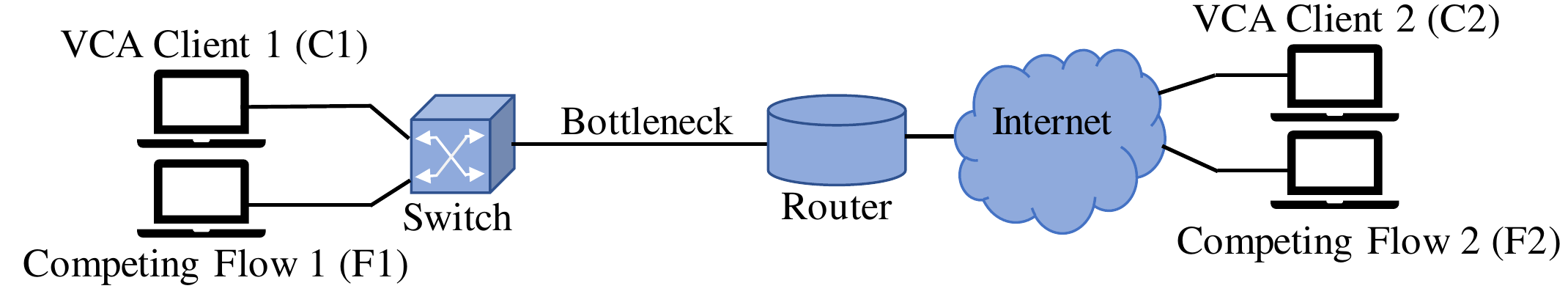}
    \caption{Setup for competition experiments.}
    \label{fig:competition-setup}
\end{figure}

\begin{figure*}[t!]
    \centering
    \begin{subfigure}[t]{.33\textwidth}
        \centering
        \includegraphics[width=1\textwidth]{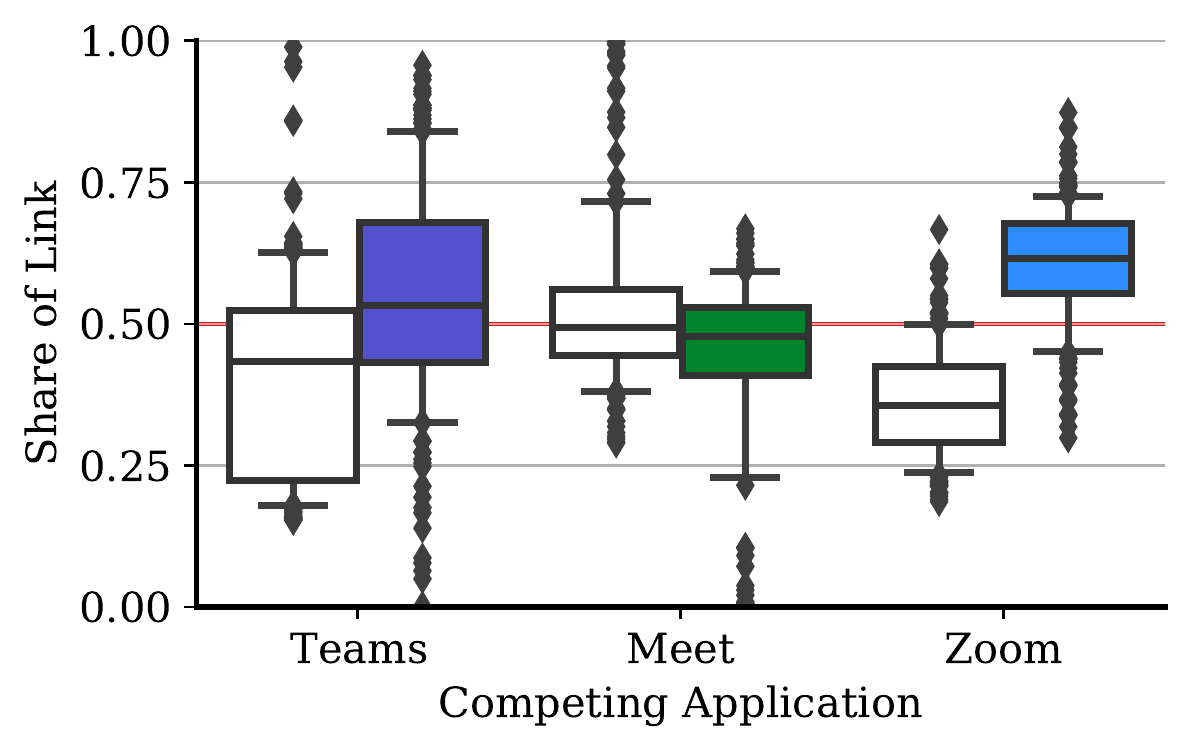}
        \caption{Incumbent: Meet}
        \label{fig:meet_ul_box}
    \end{subfigure}\hfill
    \begin{subfigure}[t]{.33\textwidth}
        \centering
        \includegraphics[width=1\textwidth]{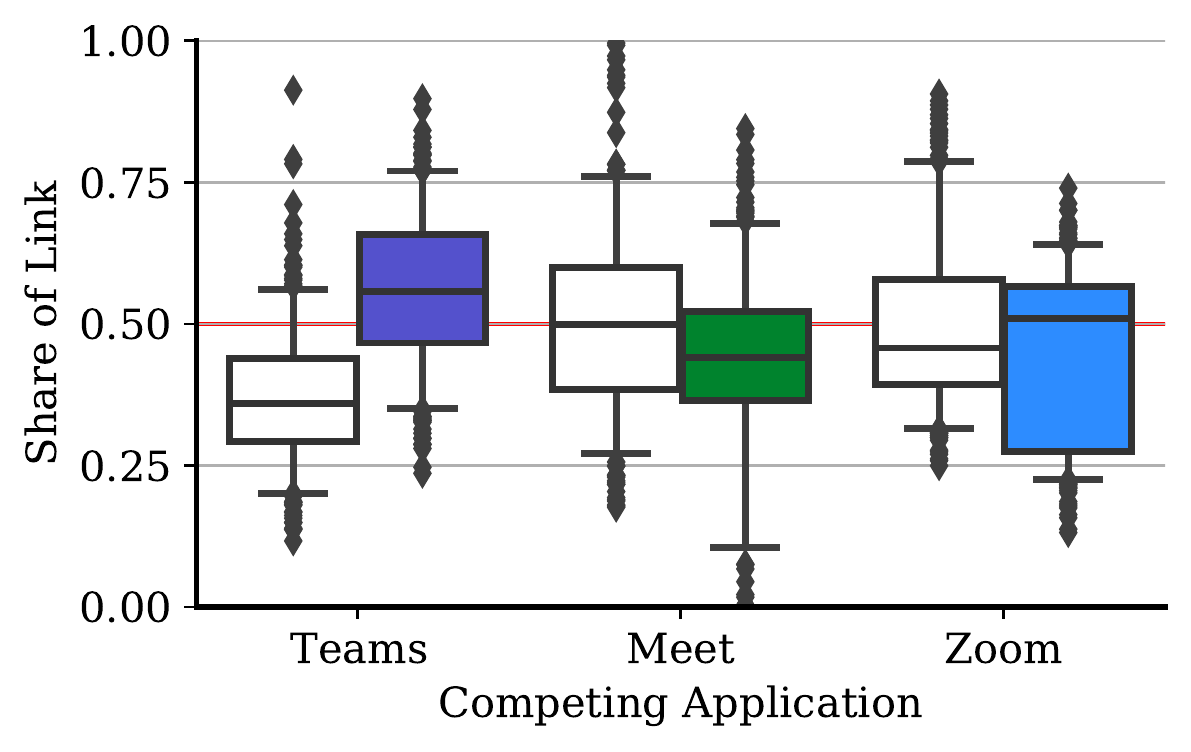}
        \caption{Incumbent: Teams}
        \label{subfig:teams_ul_box}
    \end{subfigure}
    \begin{subfigure}[t]{.33\textwidth}
        \centering
        \includegraphics[width=1\textwidth]{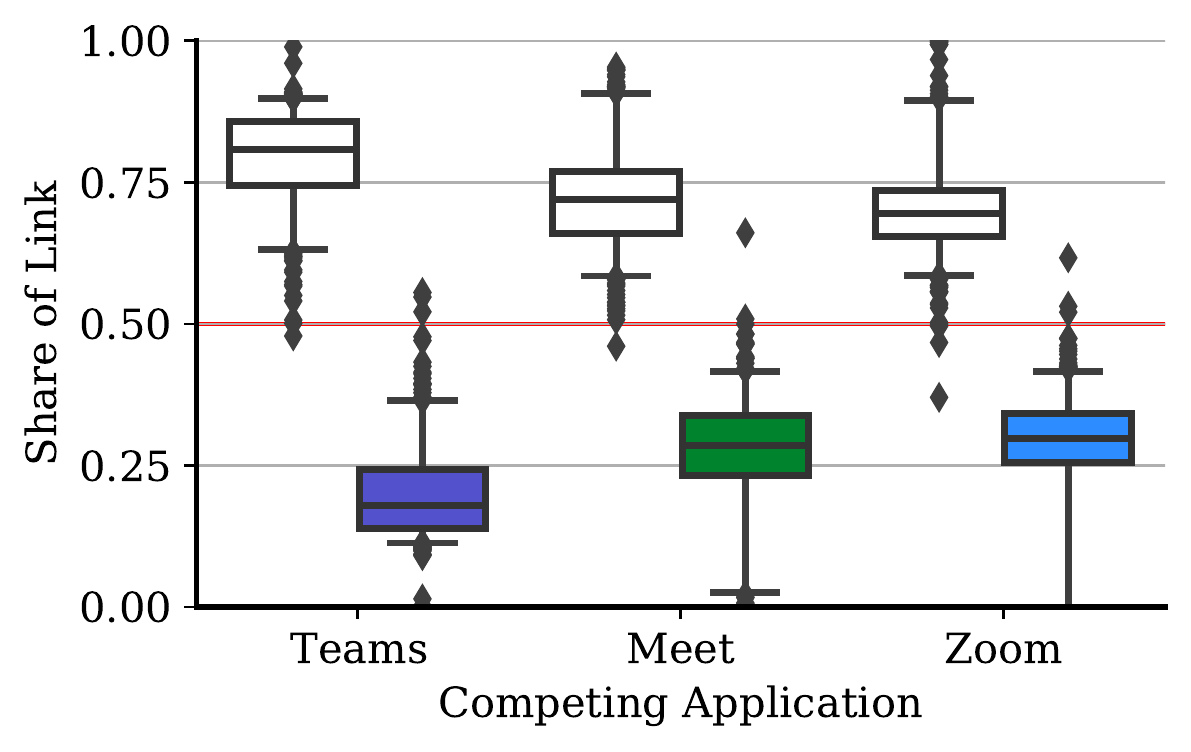}
        \caption{Incumbent: Zoom}
        \label{fig:zoom_ul_box}
    \end{subfigure}
    \caption{Share of uplink capacity used by VCAs in competition with VCAs on a 0.5~Mbps symmetric link. White box is the incumbent application}
    \label{fig:boxplot-upld}
\end{figure*}

While the intermediary server for Zoom and Meet performs congestion control,
in the case of Teams, this server acts only as a relay. During a
Teams call, C2 will recognize C1 has limited downstream capacity and 
send at only the bitrate it knows C1 can receive.
Once C1 has more available bandwidth, C2 must discover this: it must first
probe the connection before returning to its average sending rate.
Figure~\ref{fig:interrupt-sender} illustrates how C2's sending rate to the intermediary server does not
change during a Meet call, but drops below C2's downlink capacity during a
Teams call, leading to the slow recovery.

Zoom's downstream and upstream bitrates remain at the level of available
capacity during the disruption; Meet and Teams suffer more
serious degradation. Zoom's efficient network utilization and response to
disruption may be due to its encoding mechanisms, which allows it to  
effectively control the encoding parameters (i.e., SVC-encoded layers, FPS,
resolution etc.) to match the target bitrate.  
\vspace{5pt}
\begin{mdframed}[roundcorner=5pt, backgroundcolor=black!10] \noindent
    \textbf{Takeaways}: Modern VCAs are slow to recover from reductions in uplink 
    capacity, with all three requiring more than 25 seconds to recover from severe
    interruptions to 0.25 Mbps. Only Teams is consistently slow to recover
    from disruptions to downlink capacity, even showing degradation in
    response to more moderate capacity reductions (e.g., to 1 Mbps).
    This result can be attributed to each VCA's media sending mechanism and how
    congestion control is implemented at the intermediate server for each VCA. 
\end{mdframed}

\section{Competing Applications}
\label{sec:competition}

With work and school continuing online, it is even more common to have multiple
applications simultaneously using the home network, potentially leading to
competition between any combination of VCAs, video streaming applications, or other popular
applications. In this section, we measure how VCAs perform in the presence of
other applications sharing the same bottleneck link. We focus on
link sharing with other VCAs, a single TCP flow (iPerf3), and two popular
video streaming applications, Netflix and YouTube (which uses QUIC).

\begin{figure}[t!]
\centering
\begin{subfigure}[t]{.4\textwidth}
    \centering
    \includegraphics[width=1\textwidth]{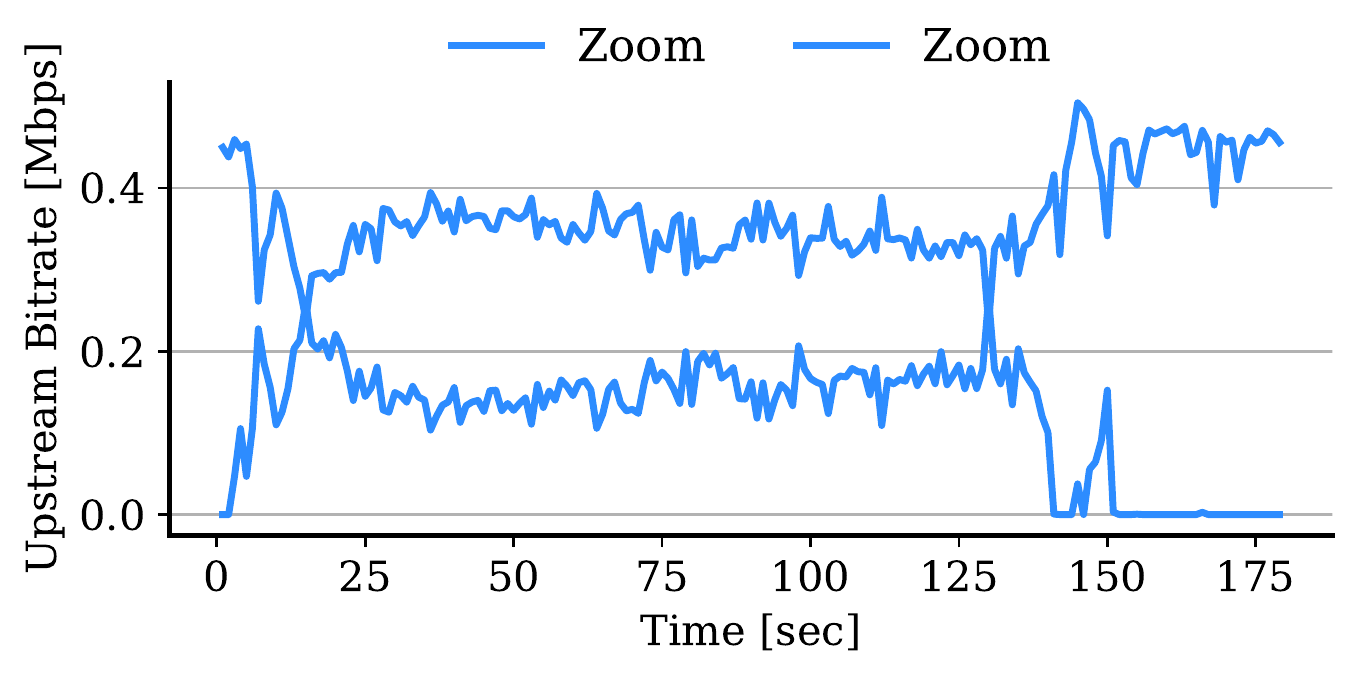}
    \caption{Zoom vs. Zoom}
    \label{subfig:zoom_zoom_0_5}
\end{subfigure}\hfill
\begin{subfigure}[t]{.4\textwidth}
    \centering
    \includegraphics[width=1\textwidth]{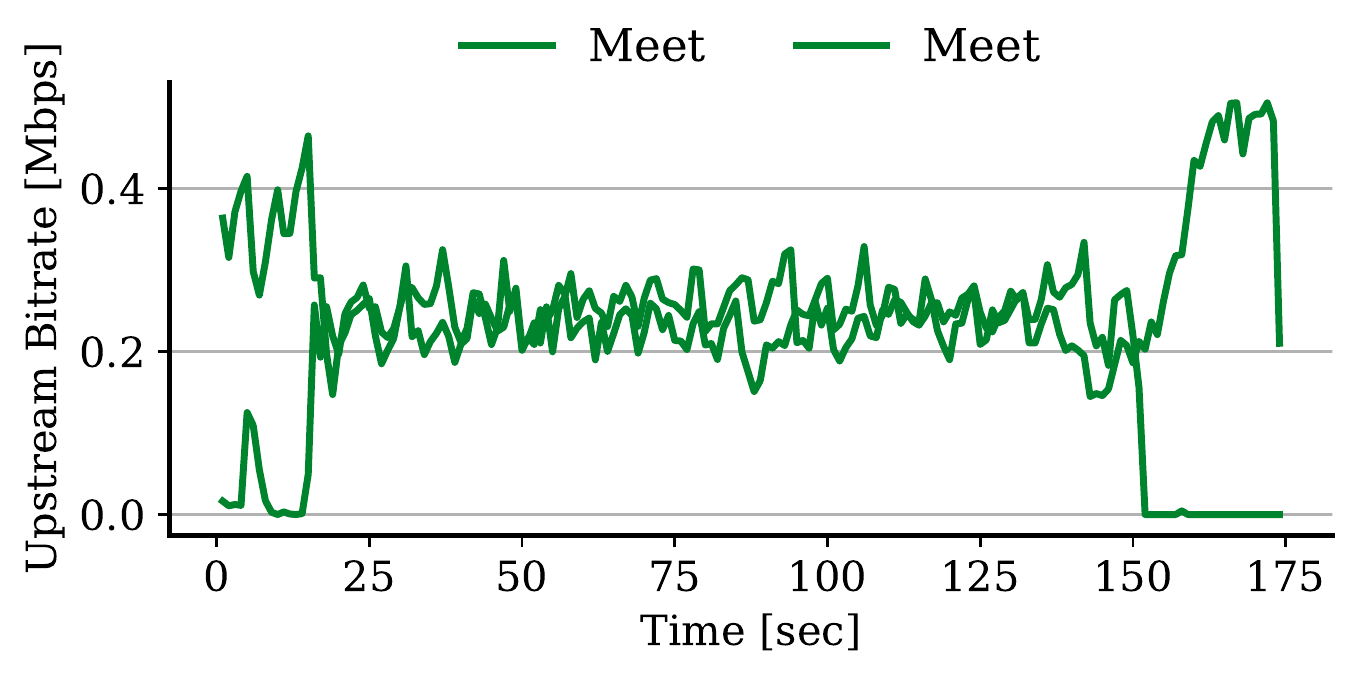}
    \caption{Meet vs. Meet}
    \label{subfig:meet_meet_0_5}
\end{subfigure}
\caption{Upstream bitrate of VCAs in competition on a 0.5~Mbps capacity link. Upper trendline is incumbent application and lower is competing application.}
\label{fig:meet-zoom-upld-0.5}
\end{figure}

\begin{figure*}[t!]
\centering
\begin{subfigure}[t]{.33\textwidth}
    \centering
    \includegraphics[width=1\textwidth]{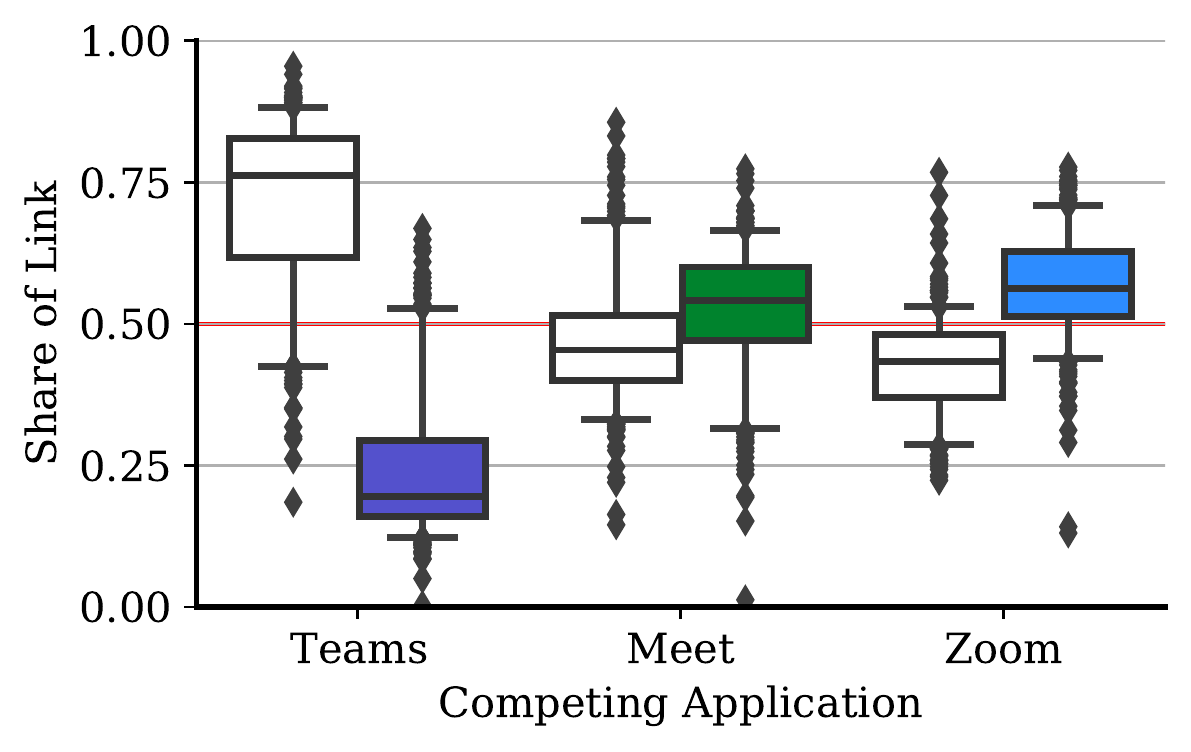}
    \caption{Incumbent: Meet}
    \label{fig:meet-dl-boxplot-0.5}
\end{subfigure}\hfill
\begin{subfigure}[t]{.33\textwidth}
    \centering
    \includegraphics[width=1\textwidth]{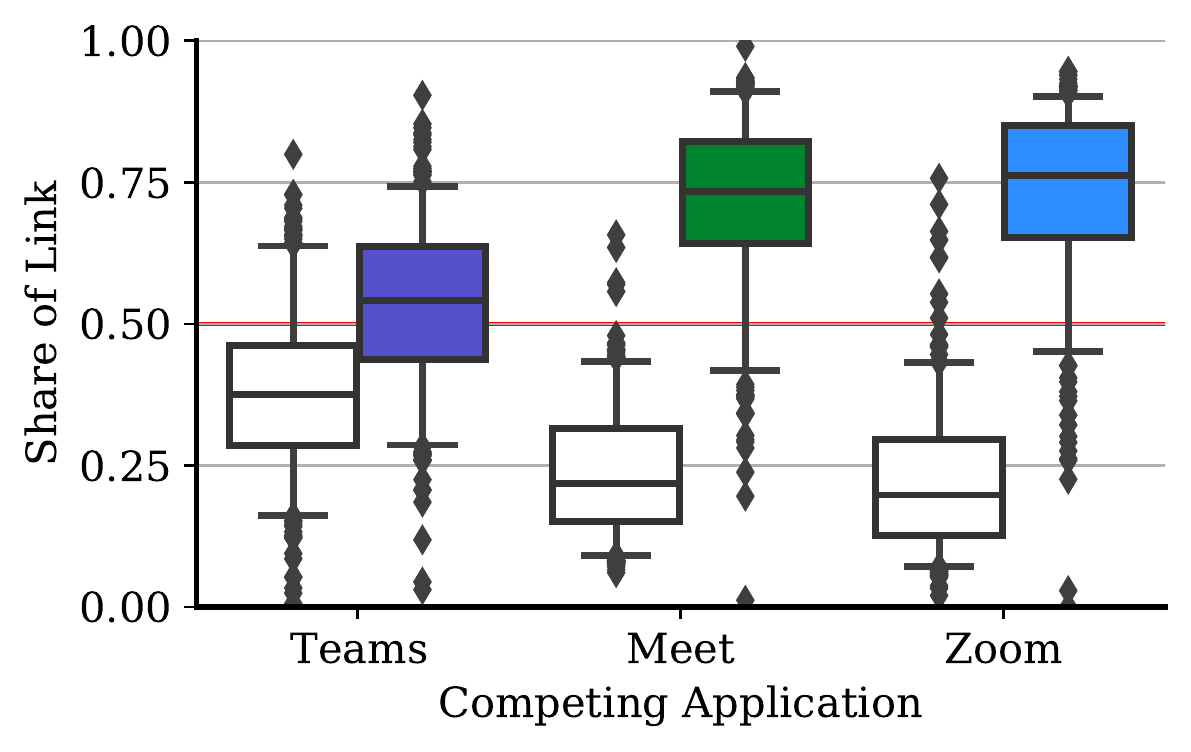}
    \caption{Incumbent: Teams}
    \label{fig:teams-dl-boxplot-0.5}
\end{subfigure}
\begin{subfigure}[t]{.33\textwidth}
    \centering
    \includegraphics[width=1\textwidth]{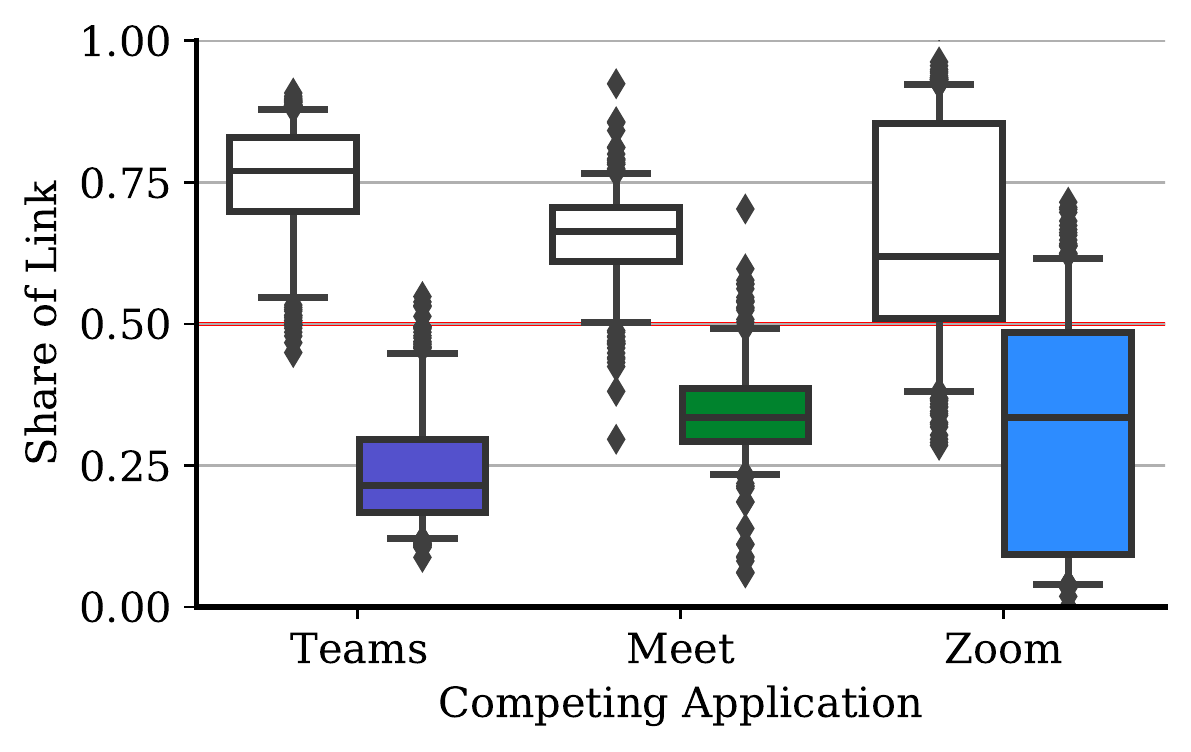}
    \caption{Incumbent: Zoom}
    \label{fig:zoom-dl-boxplot-0.5}
\end{subfigure}
\caption{Share of downlink capacity used by VCAs in competition with other VCAs on a 0.5~Mbps symmetric link. White box is the incumbent application.}
\label{fig:dnld-boxplot}
\end{figure*}

\begin{figure}[t!]
\centering
\begin{subfigure}[t]{.5\textwidth}
    \centering
    \includegraphics[width=0.8\textwidth]{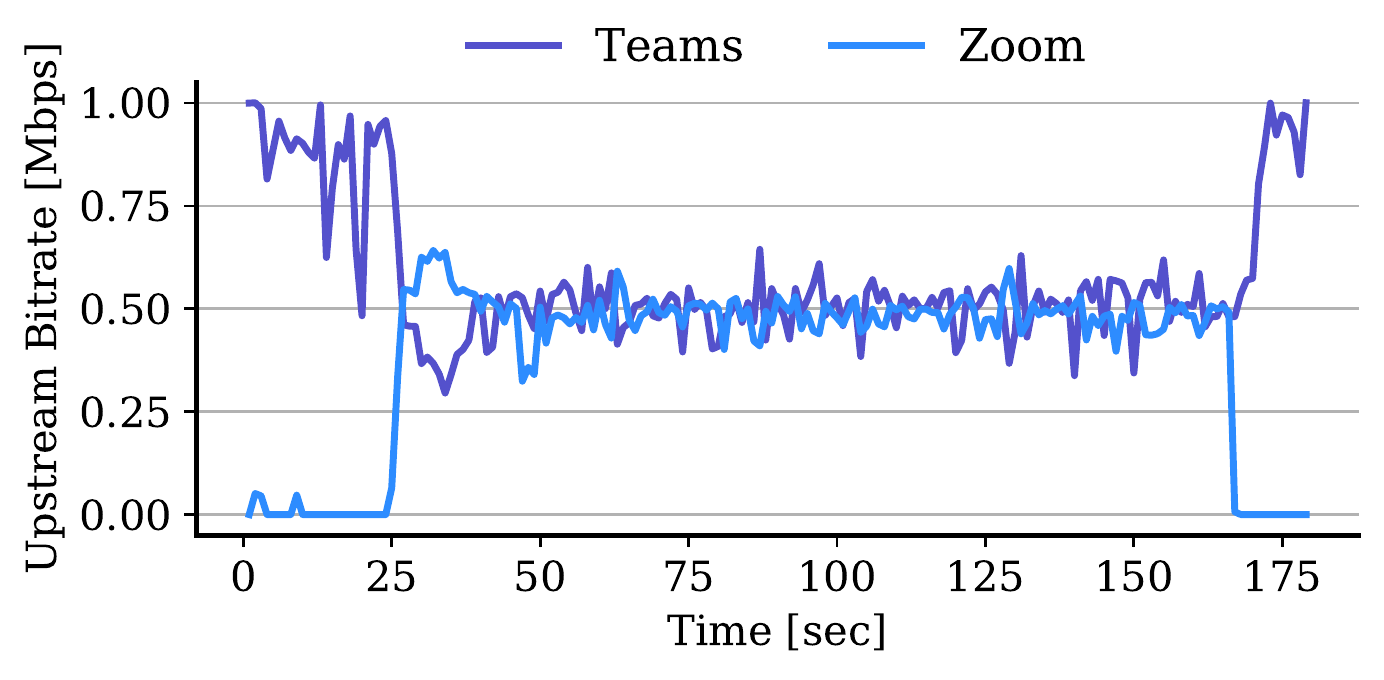}
    \caption{Uplink}
    \label{fig:teams-zoom-up-1}
\end{subfigure}\hfill
\begin{subfigure}[t]{.5\textwidth}
    \centering
    \includegraphics[width=0.8\textwidth]{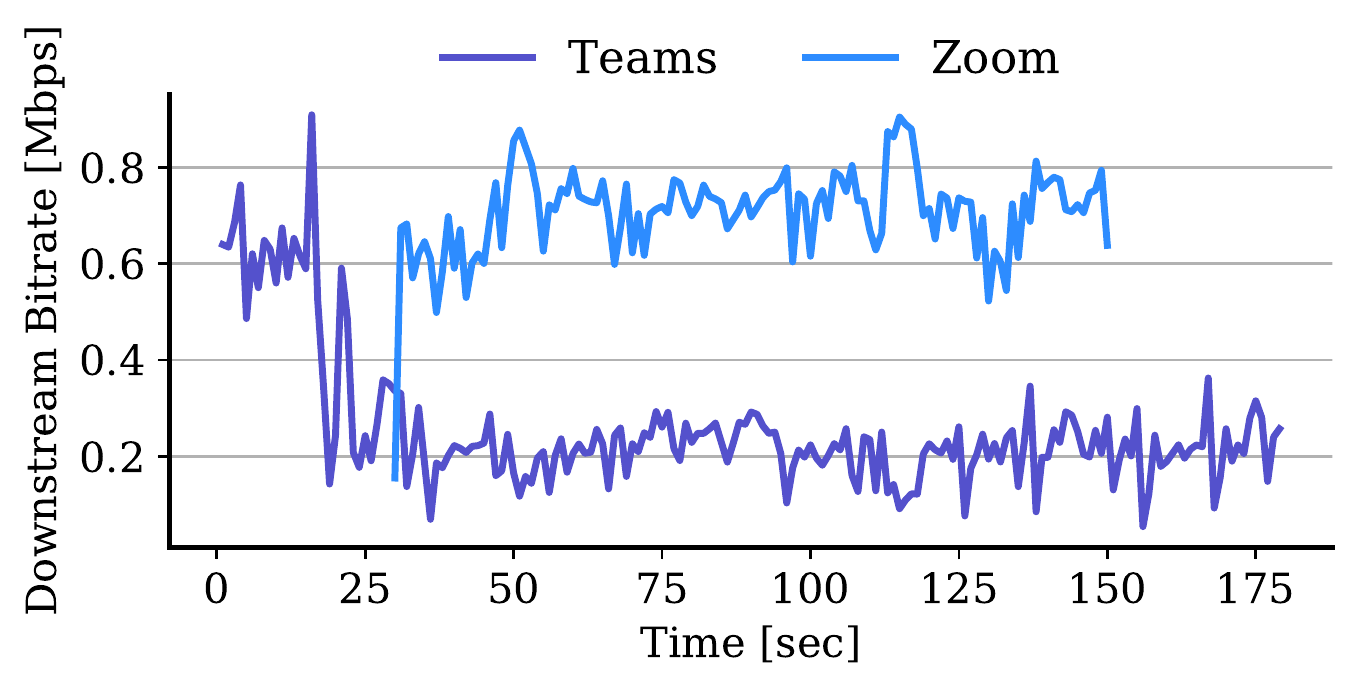}
    \caption{Downlink}
    \label{fig:teams-zoom-down-1}
\end{subfigure}
\caption{Comparison of Teams competition behavior for uplink and downlink on a 1~Mbps symmetric link. Teams is the incumbent application.}
\label{fig:teams-zoom-1}
\end{figure}

\noindent \textbf{Method}: As illustrated in Figure \ref{fig:competition-setup}, the setup for these experiments differs slightly from earlier ones. 
Instead of connecting the client directly to the router, the two matched clients, C1 and F1, are connected to the router via a switch. 
The link capacity between the switch and the router is set on the router. For each test, C1 (the incumbent application) first establishes a VCA call with C2.
Approximately 30 seconds later, F1 initiates the competing application, which lasts for two minutes.
F1's counter-party, F2, depends on the type of competing application.
If the competing application is a VCA call, then F2 is another laptop.
If the application is an iPerf3\footnote{The iPerf3 server uses TCP CUBIC and is  within the same network (average RTT ~2ms).} (TCP) flow, then F2 is a server on the same network.
  If F1 is Netflix or Youtube (launched on Chrome), then F2 is a server for the respective service. 
After the competing application terminates, the incumbent application continues for an additional minute.
We repeat each experiment 3 times, with link capacity varied symmetrically at \{0.5, 1, 2, 3, 4, 5\}~Mbps.

\subsection{VCA vs. VCA}

VCAs can achieve their nominal bitrate when the link capacity is
4~Mbps or greater, which is expected because the link capacity exceeds the sum of
nominal bitrates of each VCAs. We then focus on the operating regime when the link
capacity is lower, leading to competition between the VCAs. 
With only two competing clients, we use the proportion of the link shared as a
metric to assess fairness. We call a VCA ``aggressive'' if it uses more than
half of the link capacity under competition, and ``passive'' otherwise.

Considering the upstream direction, we observe differences in how each VCA shares the
link. Figure~\ref{fig:boxplot-upld} shows a box plot of the upstream share of
an incumbent VCA and a competing VCA when the uplink capacity is 0.5~Mbps.  We
find that an incumbent Meet client shares the link fairly with a new Meet or
Teams client but backs off when a Zoom client joins (see
Figure~\ref{fig:meet_ul_box}). The results are similar for Teams except it
receives a slightly higher share while competing with Zoom compared to Meet.
Interestingly, Zoom is highly aggressive, both as an incumbent and a competing
application, using at least $75\%$ of the link capacity in the case it is an
incumbent client (see Figure~\ref{fig:zoom_ul_box}). In fact, Zoom's
congestion control is not even fair to itself.

Figure~\ref{subfig:zoom_zoom_0_5} illustrates this effect more clearly, with
the link sharing between two Zoom clients for a single experiment under
0.5~Mbps uplink capacity. We contrast this with the link sharing between two
Meet clients in Figure~\ref{subfig:meet_meet_0_5}, where both sessions converge to their
fair share of 0.25~Mbps. The results are similar for other uplink capacities,
with aggressive applications leaving more room for new clients if they achieve
their nominal bitrate.  

We find similar results for Zoom and Meet when downstream capacity is
constrained.
However, we find that Teams is passive when sharing downstream capacity, backing off to
all other VCAs, including other Teams flows. Figure~\ref{fig:teams-dl-boxplot-0.5} shows the
link share of an incumbent Teams client compared to the competing VCA under
0.5~Mbps downlink capacity. Teams achieves only 20\% of the link
when sharing with Meet and Zoom. We observe similar behavior for other downlink
capacities. Figure~\ref{fig:teams-zoom-down-1} illustrates how an incumbent
Teams client shares the link with a Zoom client at 1~Mbps downlink capacity.
Clearly the Teams client backs off to 0.2~Mbps. In contrast 
in the upstream direction, both Teams and Zoom converge to a near fair
share of the link, as shown in Figure~\ref{fig:teams-zoom-up-1}. 
Additionally, the downstream throughput in Figure~\ref{fig:teams-zoom-down-1} for Teams
degrades even before the Zoom call starts, likely because the
competing client opens the Zoom landing page to initiate a call. This can lead
to competing TCP traffic on the link and as we find in the next subsection,
Teams is extremely passive when competing against TCP.


\begin{figure}[t!]
\centering
\begin{subfigure}[t]{.5\textwidth}
    \centering
    \includegraphics[width=0.7\textwidth]{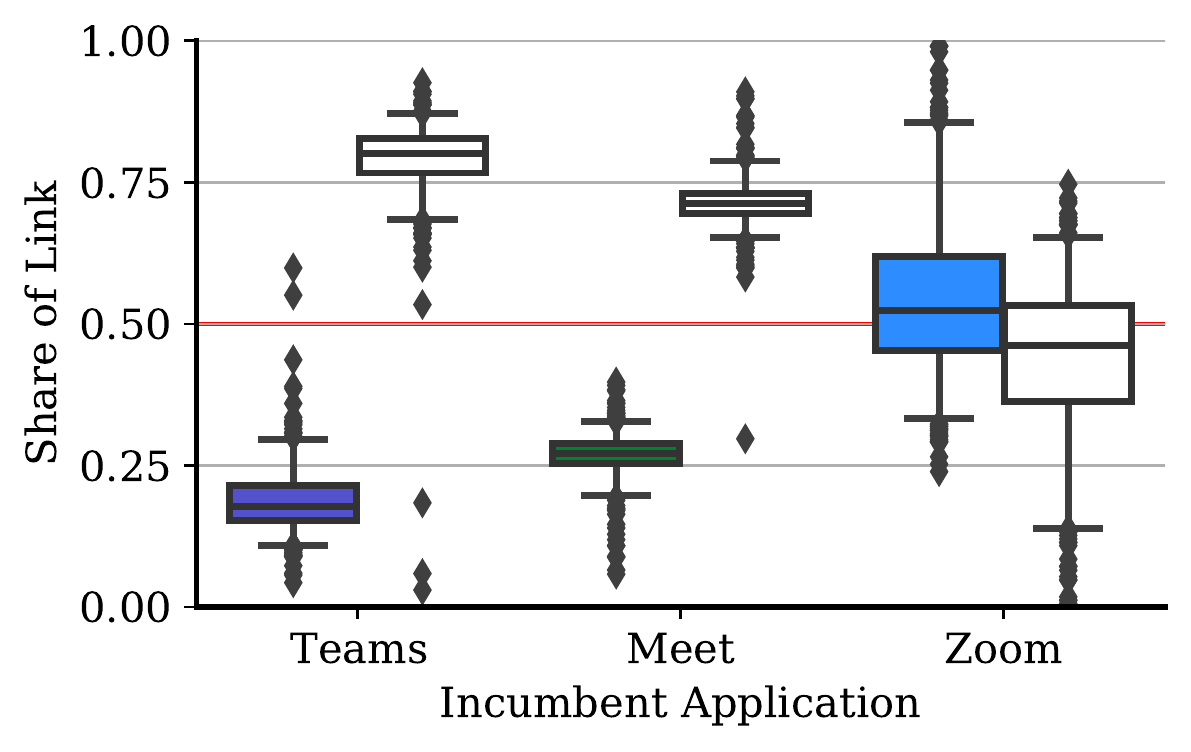}
    \caption{Downlink}
    \label{subfig:boxplot-iperf-dl}
\end{subfigure}\hfill
\begin{subfigure}[t]{.5\textwidth}
    \centering
    \includegraphics[width=0.7\textwidth]{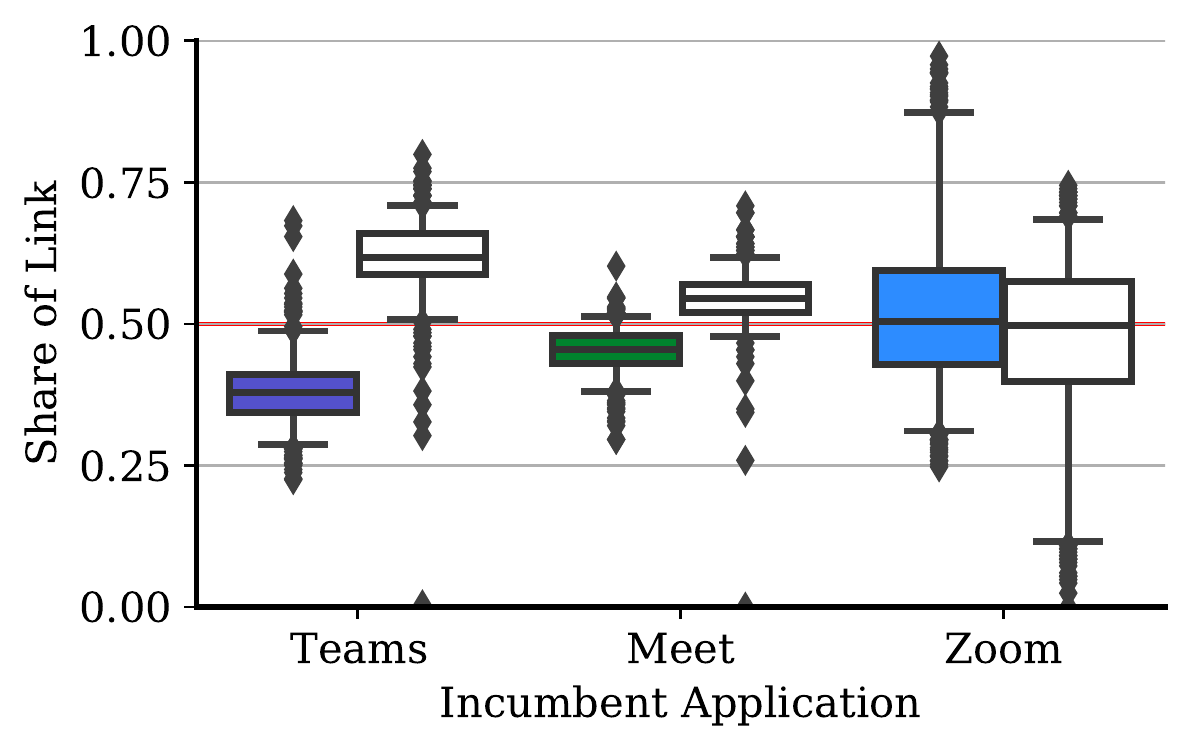}
    \caption{Uplink}
    \label{subfig:boxplot-iperf-ul}
\end{subfigure}
\caption{iPerf3 link sharing with VCAs on a 2~Mbps capacity link.}
\label{fig:boxplot-iperf}
\end{figure}

\begin{figure}[th]
    \centering
    \includegraphics[width=\linewidth]{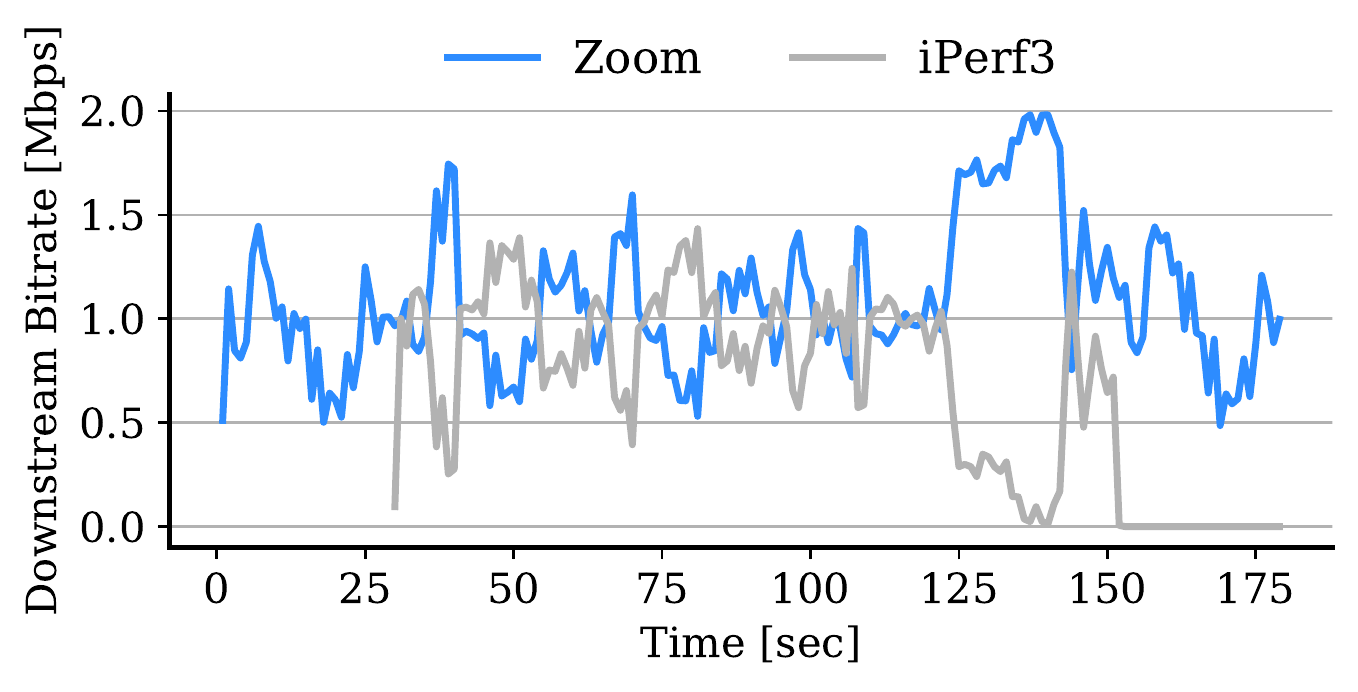}
    \caption{Example of how Zoom probing can adversely affect competing applications}
	\label{fig:zoom-iperf-dl-2}
\end{figure}

\subsection{VCA vs. TCP}

We now compare how the three VCAs compete with a 120 seconds long TCP flow
from iPerf3. In a home network, a long TCP flow can be created by file
download or upload. We find that Zoom is highly aggressive,  especially at low
uplink and downlink capacity, consuming more than 75\% of the bandwidth under
a symmetric 0.5~Mbps link. Meet is TCP-friendly in the uplink direction but
not in the downlink, consuming $75\%$ of the bandwidth at 0.5~Mbps downlink
(figures omitted due to lack of space). Teams, on the other hand, is passive
and backs-off against a TCP flow in both upstream and downstream even at high
link capacity. Figure~\ref{subfig:boxplot-iperf-ul}
and~\ref{subfig:boxplot-iperf-dl} show how each VCA shares a 2~Mbps downlink
and uplink with iPerf3, respectively. Teams is able to use 37\% in uplink and
only 20\% in the downlink even at 2~Mbps. The low upstream throughput for
Teams with a TCP flow is particularly surprising as it was able to achieve its
fair-share when competing against (more aggressive) Meet and Zoom, as shown in Figure~\ref{subfig:teams_ul_box}. At 2 Mbps, both Meet and Teams achieve their nominal bitrate allowing iPerf3 to use rest of the link. Clearly, all the VCAs are not TCP CUBIC-friendly with Meet and Zoom being highly aggressive and Teams being highly passive.

\paragraph{Anomalous Zoom Bursts}: In the previous section, Figures \ref{fig:ts_upld} and \ref{fig:ts-dnld} showed how Zoom can send bursts of data for an extended period of time following a network disruption. Figure \ref{fig:zoom-iperf-dl-2} shows how this behavior is also exhibited when in competition with iPerf3. At about 125 seconds, Zoom's increased sending rate causes iPerf3 to abruptly lower its utilization. This confirms that the temporary bursts in Zoom, likely for inferring available bandwidth, can adversely affect other applications on a scarce link.

\begin{figure}[t!]
\centering
\begin{subfigure}[t]{.35\textwidth}
    \centering
    \includegraphics[width=1\textwidth]{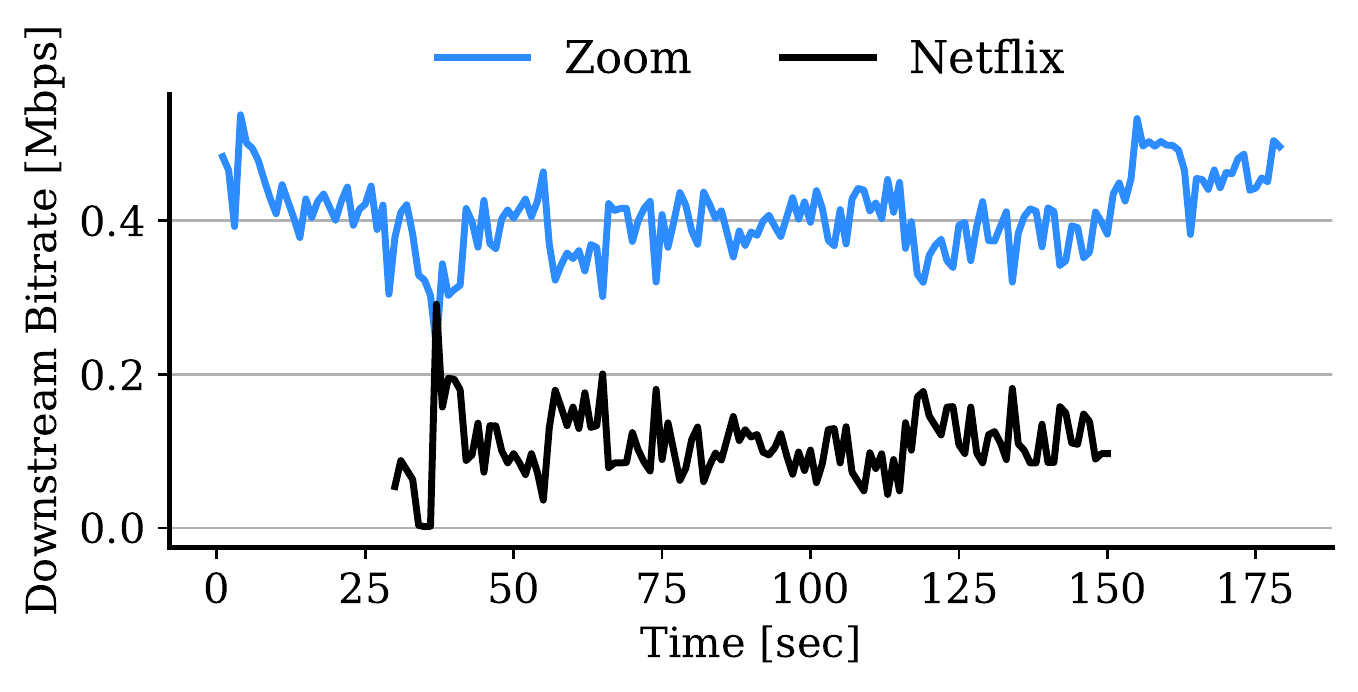}
    \caption{Downstream bitrate for Zoom and Netflix}
    \label{subfig:comp_zoom_netflix_bitrate}
\end{subfigure}\hfill
\begin{subfigure}[t]{.35\textwidth}
    \centering
    \includegraphics[width=1\textwidth]{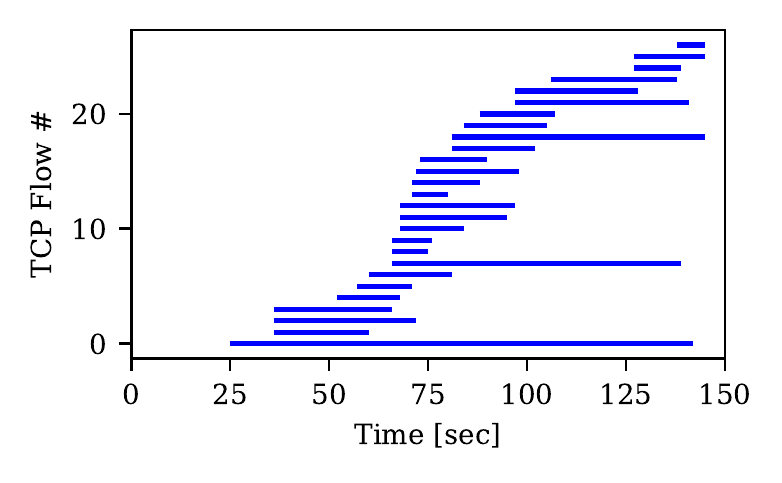}
    \caption{TCP connections opened by Netflix}
    \label{subfig:comp_netflix_conn}
\end{subfigure}
\caption{Netflix and Zoom competition on a 0.5~Mbps capacity link.}
\label{fig:comp_netflix_zoom}
\end{figure}

\subsection{VCA vs. Video Streaming}

We also compared each VCA's link share against two video streaming
applications, Netflix and YouTube,
which consume significant downstream bandwidth. 
YouTube uses QUIC, a UDP-based transport protocol,
which can be TCP-friendly depending on some configuration values~\cite{corbel2019assessing}. 
Both Meet and Zoom are
very aggressive when competing against video streaming applications, using
over $75\%$ of the link capacity. In contrast, Teams uses
less than $25\%$ while competing with YouTube and Netflix at 0.5 Mbps.  
Figure~\ref{subfig:comp_zoom_netflix_bitrate} shows this effect, when a
Netflix client competes with an incumbent Zoom client at 0.5~Mbps downstream
capacity. Zoom achieves an average throughput around 0.4~Mbps while
Netflix struggles to reach more than 0.1 Mbps. We observe this effect despite
the fact that Netflix is
known to using multiple TCP connections, especially when capacity is limited.
Figure~\ref{subfig:comp_netflix_conn} shows the TCP connections
opened by Netflix during the 120-second experiment. In total, Netflix opens 28
TCP connections, each with more than than 100 Kbits of data; at one point, it
opens 11 parallel TCP
connections. Despite this behavior, opening multiple TCP connections
does not improve link sharing for Netflix while competing with Zoom.

\begin{figure*}[tb!]
\begin{subfigure}[t]{.33\textwidth}
  \centering
   \captionsetup{width=.9\linewidth}
    \includegraphics[width=1\textwidth,keepaspectratio]{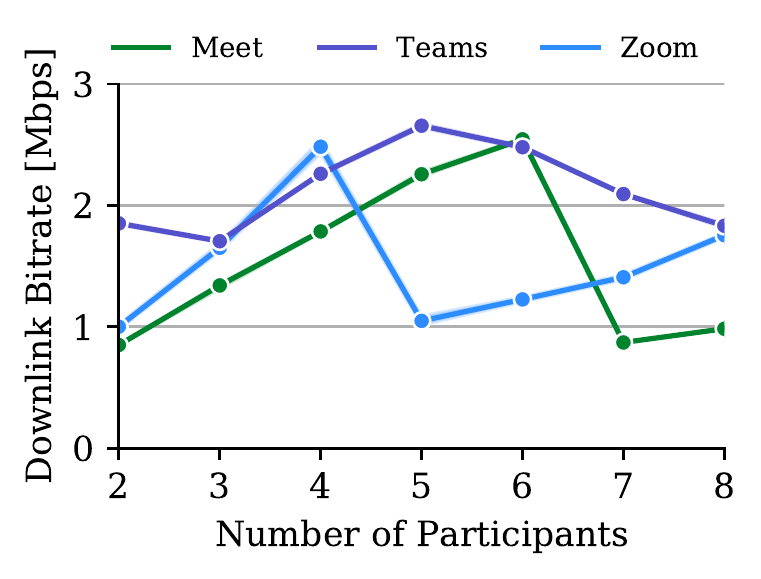}
    \caption{Downlink traffic of client whose video is viewed in gallery mode}
    \label{fig:gallery-recv}
\end{subfigure}
\hfill
\begin{subfigure}[t]{.33\textwidth}
  \centering
   \captionsetup{width=.9\linewidth}
    \includegraphics[width=1\textwidth,keepaspectratio]{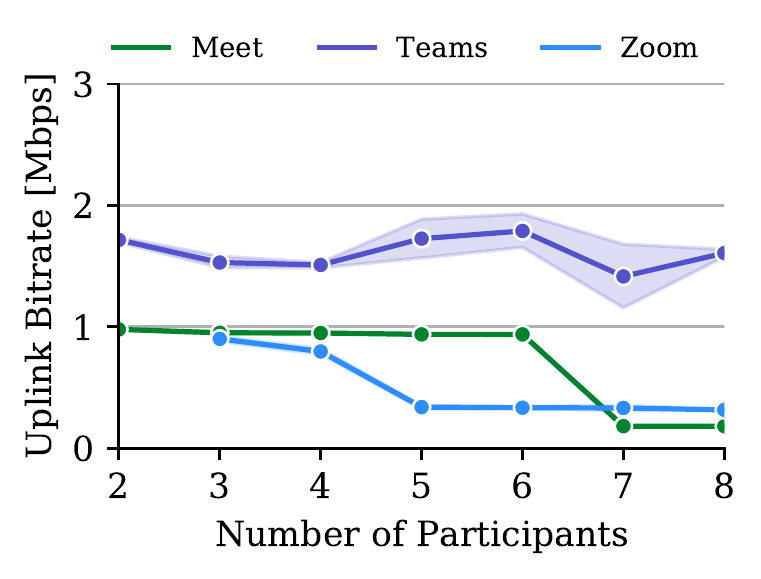}
    \caption{Uplink traffic of client whose video is viewed in gallery mode}
    \label{fig:gallery-send}
\end{subfigure}
\hfill
\begin{subfigure}[t]{.33\textwidth}
  \centering
   \captionsetup{width=.9\linewidth}
    \includegraphics[width=1\textwidth,keepaspectratio]{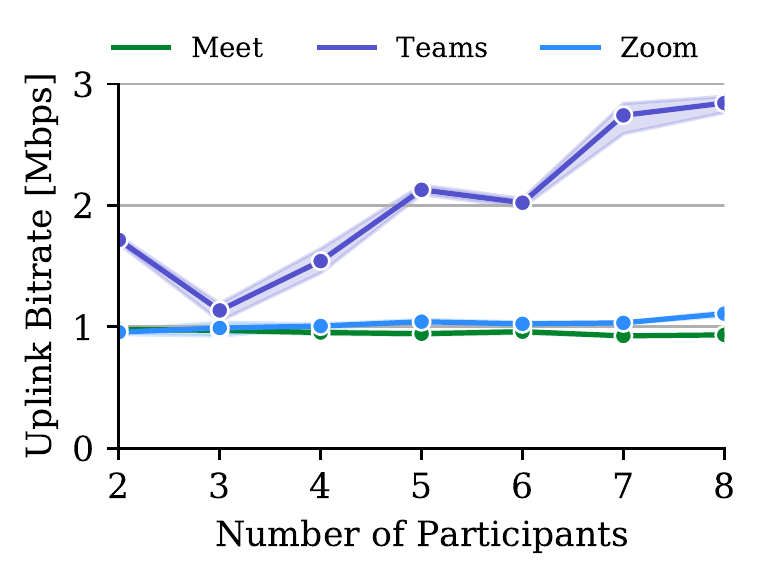}
    \caption{Uplink traffic of client whose video is pinned by all other participants (speaker mode)}
    \label{fig:speaker-send}
\end{subfigure}
\caption{Network utilization in different viewing modes. The bands represent 90\% confidence intervals.}
\label{fig:viewing-mode}
\end{figure*}

\section{Call Modalities}\label{sec:usage_modality}

People now increasingly rely on video conferencing for remote work,
school, and even large virtual conferences (such as this one!),
particularly during the COVID-19 pandemic, which has shifted many in-person
activities to virtual forums. These settings involve many
participants, which raises questions concerning performance and network
utilization of VCAs in settings involving multiple users.
We study network utilization under two prominent call modalities: 
\begin{enumerate}
    \itemsep=-1pt
    \item the
    number of participants in a call and 
\item the viewing mode.
\end{enumerate}
\noindent
We consider two
viewing modes that are common across all three VCAs: \textit{speaker} mode
wherein a specific user's video is pinned on the call and \textit{gallery}
mode, in which all participants' video are shown on the screen.  Each
experiment consists of a 2-minute call with $n$ users and specific viewing
mode. We vary the number of clients in the call from two to eight, across both
gallery and speaker modes.
We perform five experiment for each (number, viewing mode) combination; we log the
network utilization of Client C1 for each call.  Although it is
difficult to evaluate these applications for hundreds of simultaneous users,
we can nevertheless explore trends in utilization and performance for a
smaller number of users and observe various utilization trends. Larger-scale
experiments could be a possible area for future work.

\subsection{Number of Users}

To explore the effect of the number of users on utilization, we fix the
viewing mode to \textit{gallery}, which is the default viewing mode in all of
the VCAs. Figures~\ref{fig:gallery-recv} and~\ref{fig:gallery-send} show the
average downstream and upstream network utilization respectively for different
numbers of participants.  Total downstream utilization depends on both the
number of video streams and the data rate of each stream.  Utilization in both
directions typically {\em decreases} as the number of participants increases
in a \meet or \zoom call. In the case of \zoom, the uplink utilization drops
from 0.8~Mbps to 0.4~Mbps as number of participants increases from 4 to 5. For \meet, the reduction happens at n = 7, from
1~Mbps to 0.2~Mbps.  Both \meet and \zoom have tiled-screen display with each
user displayed in a separate tile.  As the number of users increases, the tile
size shrinks to accommodate all the users on the fixed size screen. For instance, \zoom uses a 2$\times$2 grid for 4 participants; switching to 5 participants creates a third row of video feeds making each individual video feed smaller.  The
sender uses this opportunity to reduce the resolution of transmitted video,
leading to reduction in upstream utilization.

We observe a similar reduction in downstream utilization at 5 and 7
participants for \zoom and \meet, respectively
(Figure~\ref{fig:gallery-recv}). However, there are also notable differences
when compared to the uplink utilization. For instance, the downlink
utilization for Google Meet increases from 1.25 Mbps to 2.5 Mbps when the
number of participants increases from 3 to 6, while upstream utilization stays
mostly constant. The trend is similar for \zoom as the number of participants
increases beyond than 5. 

\teams does not exhibit these trends: upstream utilization remains almost
constant as the number of participants changes. Downstream utilization
increases until five participants and drops as more participants join the call.
\teams has a fixed 4-tile layout on Linux. It thus displays
only a {\em subset} of participants if a call has more than four participants
which
may explain why the sending rate does not change, particularly as the number
of videos and the frame size may not
change significantly with more users. It is, however, not clear why the
downstream utilization decreases in calls with more than participants; this
phenomenon deserves more exploration. 

\subsection{Viewing Mode}

For all three VCAs, viewing a user's video in speaker mode leads to greater
uplink consumption on {\em that} user's network as compared to gallery mode.
Putting C1 on the speaker mode enlarges its tile size on other users' screen.
The sender streams the video at a high resolution to provide a high better user experience,
thus leading to increase in uplink utilization compared to when C1
was not pinned by other users. \zoom and \meet consistently send at 1 Mbps
when all clients pin C1's video, regardless of the number of participants.
Note that, only one client needed to put C1 on speaker for this behavior. Each
client can decide to pin any client independently of others.

The behavior of \teams differs from \meet and \zoom in this regard, as well.
C1's uplink utilization continues to increase from 1.25 Mbps with 3
participants to 2.9 Mbps participants when it is put on speaker mode. We
checked if the increase could be attributed to \teams communicating with
multiple destinations (e.g., with each user separately). However, we observe
that all of the traffic was directed to a single server. It is not clear what
contributes to the increase in traffic for \teams but this clearly leads to
inefficient network utilization, especially when compared to \zoom and \meet. 
\vspace{5pt}
\begin{mdframed}[roundcorner=5pt, backgroundcolor=black!10]
    \paragraph{Takeaways}: Each participant's video layout affects client's
    network utilization. Calls with more participants can {\em decrease} upstream and downstream utilization 
    for each participant, depending on how the VCA displays participant video.
    The settings of one participant (e.g., pinning a video, using speaker mode
    vs. gallery mode) can also affect the upstream utilization of {\em other}
    participants.
\end{mdframed}

\section{Related Work}\label{sec:related}

\paragraph{VCA measurement}: Some of the early VCA measurement work has focused on uncovering the design of the Skype focusing on its streaming protocols~\cite{baset2004analysis}, architecture~\cite{guha2005experimental}, traffic characterization~\cite{bonfiglio2008tracking}, and application performance~\cite{hossfeld2008analysis}. More recent work has included other VCAs and streaming contexts~\cite{xu2012video, yu2014can, azfar2016android}. Xu et al.~\cite{xu2012video} use controlled experiments to study the design and performance of Google+, iChat, and Skype. The work is further extended to include performance of the three services on mobile video calls~\cite{yu2014can}. 

Closest to our work is work by Jansen et al.~\cite{jansen2018performance} and Nistico et. al~\cite{nistico2020comparative}. Jansen et al. evaluate WebRTC performance using their custom VCA under controlled network conditions~\cite{jansen2018performance}. Emulating similar network conditions, we consider performance of commercial and more recent VCAs that widely used for education and work. Even between tested VCAs using WebRTC, namely \meet and \teamsbrowser, we find significant performance differences, likely due to different design parameters (e.g., codecs, default bitrates). Nistico et al.~\cite{nistico2020comparative} consider a wider range of recent VCAs, focusing on their design differences including protocols and architecture. Our work provides a complimentary performance analysis for a subset of the VCAs studied by them. We use the insights from their work to explain the differences among VCAs' network utilization and performance under similar streaming contexts. 

\paragraph{VCA congestion control}: Several congestion control algorithms have been proposed for VCAs. These algorithms rely on a variety of signals such as loss~\cite{handley2003tcp}, delay~\cite{carlucci2016analysis}, and even VCA performance metrics~\cite{singh2012rate} for rate control. For instance, Google Congestion Control~\cite{carlucci2016analysis}, also implemented in WebRTC, uses one-way delay gradient for adjusting the sender rate while SCReAM~\cite{johansson2015self} relies on both loss and delay along with TCP-like self-clocking. 
While the VCAs may use one or more of these variants, the exact implementation of the algorithm and parameter values vary and is proprietary. In this work, we study the efficacy of the VCA congestion control in the case of transient interruptions and background applications. A recent study by Sander et al. evaluates \zoom's congestion control along these dimensions~\cite{sandervideo}. Our work observes similar results for \zoom and also analyses more VCAs, including their fairness to each other and other popular internet applications, namely YouTube (QUIC-based) and Netflix (TCP-based).

\balance\section{Future Work}
\label{sec:discussion}

\paragraph{Generalizability to other VCA contexts}: While this paper focuses on
three popular video conferencing applications, the methods in this paper could
be used to measure utilization and performance of other VCAs, across different
device platforms. Our framework, using PyAutoGUI for automation, can be
applied to other VCAs. It can also work on other device platforms based on Linux,
MacOS, or Windows. Furthermore, we can include other network profiles that represent
other contexts, such as WiFi and cellular. We plan to release all of our
experiment data and code so that it others can extend this work 
to other contexts.

\paragraph{Application performance metrics}: Analyzing application performance
statistics can shed more light on the VCA behavior and user quality of
experience, as is also seen through a subset of our results using WebRTC
statistics in Section~\ref{subsec:application_performance}. It is challenging
to obtain application performance metrics for all VCAs, especially native
clients. Future research could explore the methods from related work, such as
using annotated videos~\cite{xu2012video} or network traffic
captures~\cite{dasari2018scalable} to infer application metrics.

\paragraph{Performance under other network conditions}: Other network factors
such as latency, packet loss, and jitter could affect VCA performance and
utilization.  Future work could explore the effects of these parameters.
Experiments in this paper also focus on issues in the last-mile access
network, when there could be problems anywhere along the end-to-end path.
However, emulating all kinds of impairments is challenging to accomplish
in-lab. An alternative could be to collect VCA performance metrics from a
large number of users in the real world. 

\paragraph{Further exploration of VCA design}: Some of our results reveal
unusual VCA behavior that we found difficult to explain and deserve further
exploration. Examples include: (1)~Some VCAs exhibit different behavior
depending on if the competition exists in the upstream or downstream
direction, (2)~\teams utilization is not consistent with the video layout for
different call modalities; and (3)~\meet shows unusual encoding behavior
(specifically, increased frame rates) when network capacity is very low.

\section{Conclusion}
\label{sec:conclusion}

We analyzed the network utilization and performance of three major VCAs under
diverse network conditions and usage modalities. Our results show that VCAs
vary significantly both in terms of their resource utilization and
performance, especially when capacity is constrained.  Differences in behavior
can be attributed to different VCA design parameter values, congestion control
algorithms, and media transport mechanisms. The VCA response also varies
depending on whether constraints exist on the uplink or downlink. Different
behavior often arises due to both differences in encoding strategies, as well
as how the VCAs rely on an intermediate server to deliver video streams and
adapt transmission to changing conditions. Finally, the network utilization of
each VCA can vary significantly depending on the call modality. Somewhat
counterintuitively, calls with more participants can actually reduce any one
participant's upstream utilization, because changes in the video layout on the
user screen ultimately lead to changes in the sent video resolutions. Future
work could extend this analysis to more VCAs, device platforms, and network
impairments. In general, however, performance can begin to degrade at upstream
capacities below 1.5~Mbps, and some VCAs do not compete well with other TCP
streams under constrained settings, suggesting the possible need for the FCC
to reconsider its 25/3 Mbps standard for defining ``broadband''.

\microtypesetup{protrusion=false}

\newpage
\bibliographystyle{ACM-Reference-Format}
\thispagestyle{plain}
\balance\bibliography{paper}

\end{sloppypar}
\end{document}